\documentclass[aps,twocolumn,floatfix,superscriptaddress]{revtex4}

\usepackage{graphicx}
\usepackage{lscape}
\usepackage{indentfirst}
\usepackage{latexsym}
\usepackage{multirow}
\usepackage{tabls}
\usepackage{epsfig}
\usepackage{multirow}
\usepackage{subfigure}
\usepackage{hyperref}

\usepackage{color}
\usepackage{amssymb}

\usepackage{amsfonts}
\usepackage{amsmath}
\usepackage{bm}
\usepackage{mathrsfs}

\usepackage{algorithm}
\usepackage{algpseudocode} % in place of algorithmic. Appears to conflict with revtex4 unless [H] option passed
\usepackage{setspace} % for pleasent spacing in algorithm

\newcommand{\mb}[1]{\mbox{\boldmath $#1$}}
\newcommand{\cI}{{\cal I}}
\newcommand{\be}{\begin{equation}}
\newcommand{\ee}{\end{equation}}

\newcommand{\mlam}{\mb{\lambda}}
\newcommand{\bigo}[1]{{\cal O}({#1})}

\def\lsim{\mathrel{\rlap{\lower3pt\hbox{\hskip1pt$\sim$}}
    \raise1pt\hbox{$<$}}}                % less than or approx. symbol
\def\gsim{\mathrel{\rlap{\lower3pt\hbox{\hskip1pt$\sim$}}
    \raise1pt\hbox{$>$}}}         % greater than or approx. symbol

\def\coordeq{ \, \mathrel{ \rlap{\hbox{\hskip-2.5pt$=$} }
    \raise4pt\hbox{$\cdot$}} \, }                % equal in a particular coordinate system
\begin{document}

\title{Gravitational wave parameter estimation with compressed likelihood evaluations}

\def\addUMDa{Center for Scientific Computation and Mathematical Modeling, Department of Physics, Joint Space Sciences Institute, Maryland Center for Fundamental Physics. University of Maryland, College Park, MD 20742, USA}

\def\addUMDb{Department of Physics, Joint Space Sciences Institute, Maryland Center for Fundamental Physics. University of Maryland, College Park, MD 20742, USA}

\def\addCaltech{TAPIR, MC 350-17, California Institute of Technology,
Pasadena, CA, 91125, USA}

\def\addCambridge{Institute of Astronomy, Madingley Road, Cambridge, CB30HA, United Kingdom}

\author{Priscilla Canizares}
\affiliation{\addCambridge} 

\author{Scott E. Field}
\affiliation{\addUMDb}

\author{Jonathan R. Gair}
\affiliation{\addCambridge} 

\author{Manuel Tiglio}
\affiliation{\addUMDa}
\affiliation{\addCaltech}

\begin{abstract}
One of the main bottlenecks in gravitational wave (GW) astronomy is the high cost of performing parameter estimation and GW searches on the fly. We propose a novel technique based on Reduced Order Quadratures (ROQs), an application and data-specific quadrature rule, to perform fast and accurate likelihood evaluations. These are the dominant cost in Markov chain Monte Carlo (MCMC) algorithms, which are widely employed in parameter estimation studies, and so ROQs offer a new way to accelerate GW parameter estimation.
We illustrate our approach using a four dimensional GW burst model embedded in noise. We build an ROQ for this model, and perform  four dimensional MCMC searches with both the standard and ROQs quadrature rules, showing that, for this model, the ROQ approach is around $25$ times faster than the standard approach with essentially no loss of accuracy. The speed-up from using ROQs is expected to increase for more complex GW signal models and therefore has significant potential to accelerate parameter estimation of GW sources such as compact binary coalescences.
\end{abstract}

\maketitle

%%%%%%%%%%%%%%%%%%%%%%%%%%%
\section{Motivation and context}
%%%%%%%%%%%%%%%%%%%%%%%%%%%

Computing correlations between data and  models described by  large dimensional parameter spaces is an important aspect of many scientific disciplines. Obtaining estimates of the parameters of observed signals is crucial to extract the most from multi-billion dollar experiments such as gravitational wave (GW) detectors (i.e., advanced LIGO, advanced Virgo, Indigo, and KAGRA)~\cite{LIGO_web,VIRGO_web,GEO_web,KAGRA_web}.
However, carrying out parameter estimation on large dimensional parameter spaces can be computationally expensive. Costs grow further if several different models  or alternative theories of gravity (see, for example, ~\cite{Abadie:2010cfa,Brown:2012gs,Ajith:2011ec,Centrella:2010zf,Buonanno_etal:PRD70,Buonanno:2002fy,Pan_etal:PRD69,Littenberg:2012uj} and ~\cite{Chatziioannou:2012rf,Yunes:2009ke,Canizares:2012is}, respectively) are used to analyse the data as a prelude to Bayesian model selection. It is therefore of great importance to develop efficient methods for analysing the data to ensure that all the desired science can be extracted from the data in a reasonable time.

One of the primary methods for computing the 
probability distribution for the parameters of a given signal in a data set 
is Markov chain Monte Carlo (MCMC). This requires evaluating the posterior probability of the model parameters throughout parameter space. When the likelihood and hence posterior probability is expensive to evaluate, MCMC algorithms can become computationally prohibitive. 
In such cases, approximate methods such as the Fisher matrix are widely used because they  are significantly cheaper than a full Bayesian analysis. Several rather optimistic assumptions, however, such as high signal-to-noise ratios are often not satisfied in practice. Recently, other sampling approaches \cite{Vallisneri:2011ts,Cornish:2010kf} for computing the maximum likelihood estimator have been proposed for low signal-to-noise scenarios.

An alternative way to improve the speed of MCMC algorithms is to reduce the cost of evaluating the likelihood at each parameter space point.
This strategy has motivated work on directly interpolating the likelihood \cite{Mitra:2005gm,brown_sc_2013_13,Cannon:2012gq,Smith:2012du} and training a neural network to learn likelihood data on-the-fly~\cite{graff2012bambi}. At least in the case of direct interpolation there could be technical obstacles for likelihoods which require waveforms with many cycles and/or higher dimensionality~\cite{brown_sc_2013_13,Smith:2012du}. In this paper we describe a novel technique for fast, accurate calculations of correlations between data and 
modeled waveforms, fine tuned for applications such as MCMC. The approach is based on Reduced Order Modeling (ROM) and, as such, aims to significantly reduce the problem's dimensionality by exploiting redundancies. The result is a compressed representation of the likelihood thereby reducing the cost of each evaluation. Generalizations to higher dimensions and/or many cycles are readily handled within the method's existing framework~\cite{antil2012two}.

Within typical GW physics applications, the number of required correlations quickly grows with the number $p$ of physical parameters and the number of GW cycles. For example, the number of search templates scales as $\sim (1-\mathrm{MM})^{-p/2}$ \cite{Owen:1995tm}, where $\mathrm{MM}$ is the minimal match of the catalog. For a compact binary coalescence with $p = 8$ intrinsic parameters lasting for $10^5$ cycles we could need up to $\sim 10^{40}$ templates for a fully coherent search \cite{Babak:2009ua}. In light of these scalings there is an obvious need for reducing the cost of each correlation. 

Correlation costs typically scale with the length $N$ of the data, which depends on both the observation time and sampling rate. Furthermore, standard  fast converging numerical integration rules for smooth functions, such as Gaussian quadratures, lose their fast convergence in the presence of noisy (non-smooth) data. 
In this paper we show how integrals with noisy data can be computed with a cost not set by the Nyquist sampling rate or observation time~\cite{Allen:2005fk}, but rather the ``information content" of the gravitational waveforms themselves. The integration converges fast, typically exponentially, with the number of sparse data samples $m$ drawn from the full data set, {\em even in the presence of noise}. The overall likelihood cost is thereby reduced to $m \ll N$. 

Our approach for speeding up correlation computations is based on a recently proposed Reduced Order Quadrature (ROQ) for parametrized functions \cite{antil2012two}. Reduced order quadratures combine dimensional reduction with the Empirical Interpolation Method (EIM) \cite{Barrault2004667,Maday_2009} to produce a nearly optimal quadrature rule for parametrized systems. To do so, it exploits smooth dependence with respect to parameter variation, when available, to achieve very fast convergence with the number of data samples. Even in the absence of noise, in many cases ROQs outperform the best known quadrature rule (Gaussian quadratures) for generic smooth functions~\cite{antil2012two}. The key aspect of this apparent super-optimality is to leverage information about the space of functions in which we are interested. 

In the context of GW parameter estimation, the use of ROQs can significantly improve the performance of existing numerical algorithms by reducing the computational cost of computing a waveform overlap (correlation) with the data. Here we illustrate this application of ROQs to GW parameter estimation using a simple model of a sine-Gaussian GW burst waveform. This model is chosen as a toy one to illustrate the method. Although such waveforms have been used in GW searches (see, for example, \cite{Abadie:2012rq}), the cost of their likelihood evaluations is not significant, so we are not suggesting that this application is one for which ROQs are required. However, we demonstrate that even for such a simple model the speed-up from ROQs is significant and we expect that comparable or greater speed-ups will be possible for more complex GW signal models~\cite{antil2012two}.

This paper is organised as follows. In section \ref{sec:overview} we present an overview of the proposed approach. In sections \ref{sec:ROM}, \ref{sec:EIM} and \ref{sec:ROQ} we introduce the building blocks of the method; namely, Reduced Order Modelling, the Empirical Interpolation Method and Reduced Order Quadratures, as well as the GW burst model. Finally, in Sec.~\ref{sec:MCMC_ROQ}  we apply the ROQ approach to perform a MCMC search using the burst model, explicitly showing that ROQ can considerably speed up MCMC computations. Among the new aspects that we address compared to \cite{antil2012two} are how to deal with the arrival time of the GW signal, and the application of the technique to noisy data.  In Appendices \ref{sec:RBapp} and \ref{sec:EIMapp} we summarise the greedy approach for generating a Reduced Basis, and the Empirical Interpolation Method, respectively. 

%%%%%%%%%%%%%%%%%%%%%%%%%%%
\section{Methodology} \label{sec:overview}
%%%%%%%%%%%%%%%%%%%%%%%%%%%
In this paper we are interested in improving the performance of GW parameter estimation by using ROQs. We assume that the detected data stream is given by $s(t) = h(t;\mb{\lambda}) + n(t)$, where $h(t;\mb{\lambda})$ is the GW signal that we want to characterise, which depends on a multi-dimensional set of source parameters $\mb{\lambda}$, and $n(t)$ is instrumental noise. 

In the context of Bayesian parameter estimation the posterior probability distribution function (PDF) provides complete information about the parameters of the signal:
%~~~~~~~~~~~~~
\begin{equation}\label{eq:pdf}
p\left(\mb{\lambda}|s\right) := {\cal C} p\left(\mb{\lambda}\right) P(s|\mb{\lambda})  \, .
 \end{equation}
%~~~~~~~~~~~~~
Here $p\left(\mb{\lambda}\right)$ is the prior probability density, ${\cal C}$ an overall normalization constant, and $P(s|\mb{\lambda})$ is the likelihood that the true parameter values are given by a particular $\mb{\lambda}$, or in other words, the likelihood that the signal is present in the data stream. For Gaussian, stationary noise the likelihood is
%~~~~~~~~~~~~~
\begin{equation}\label{eq:likelihood}
 P\left(s|\mb{\lambda}\right) \propto\,\text{exp}\left({-\chi^2/2}\right),  
\end{equation}
%~~~~~~~~~~~~~
where
%~~~~~~~~~~~~~
\begin{equation}\label{eq:chi}
\chi^2 := \langle n | n  \rangle = \langle s(\cdot) -h \left(\cdot;\mb{\lambda}\right)| s(\cdot) -h \left(\cdot;\mb{\lambda}\right) \rangle  
\end{equation}
%~~~~~~~~~~~~~
is the weighted norm of the noise realization $n(t)$, defined by the weighted inner product (see e.g.~\cite{Babak:2009ua}) 
%~~~~~~~~~~~~~
\begin{align}\label{eq:scalarprod}
\left<a\left|b\right.\right> 
 = 4 \Re \int_{f_\mathrm{min}}^{f_\mathrm{max}}\frac{ \tilde{a}(f) \tilde{b}^{*}(f)}{\tilde{S}_{n}(f)} df , 
\end{align}
%~~~~~~~~~~~~~
with $^{*}$ denoting complex conjugation and $\tilde{S}_{n}(f)$ the power spectral density of the detector's noise. Owing to the form of $\tilde{S}_{n}(f)$ in GW physics, the lower limit of integration in Eq.~(\ref{eq:scalarprod}) is sometimes replaced by $f_\mathrm{min} > 0$. 

When dealing with high dimensional problems, the process of mapping the likelihood (or the posterior) surface can become very expensive. MCMC algorithms are a useful technique for searching through such large spaces, by following a random walk in parameter space, with the probability of a sample being chosen at any point being proportional to the posterior probability. However, since a MCMC search depends on the number of sampling points, as well as the dimensionality of the problem, it can still be a very expensive algorithm and in many cases prohibitively so.

This paper proposes application and data-specific quadrature rules for scenarios such as GW parameter estimation, where correlations between noisy data and a family of functions have to be repeatedly evaluated. The quadrature rules employed here are a variation of the ROQ introduced in Ref.~\cite{antil2012two} for the case $n(t)=0$, and their construction follows several layers of dimensional reduction that are explained in the different sections of this paper, namely:
\begin{enumerate}
\item {\bf Construct a basis for the space of waveforms of interest. Offline stage}. 

Described in Sec.~\ref{sec:ROM}. A Reduced Basis-greedy approach has several advantages, including an approximation to the most relevant points in parameter space, but the proposed ROQ can use any choice of a ``good" basis. 
\item {\bf Identify the empirical interpolation points associated with the above basis. Offline stage}.

Described in Sec.~\ref{sec:EIM}. This step provides, through a greedy approach, the set of most relevant points in the physical dimension(s), and a nearly optimal global interpolant associated with the basis constructed in Step 1. These EIM nodes are to be used as integration points in the ROQ rule. 
\item {\bf Given any stream of data, construct the weights of the ROQ.  Startup stage}. 

Described in Sec.~\ref{sec:ROQ}. These weights are linear combinations of correlations between the data and the basis elements of Step 1.

\item{\bf Fast likelihood evaluations. Online stage}.

Described in Sec.~\ref{sec:MCMC_ROQ}. The ROQ uses the nodes computed in Step 2 and the weights computed in Step 3 to perform fast and accurate evaluations of overlaps between the data and any waveform within the model.   
\end{enumerate} 
Section \ref{sec:MCMC_ROQ} discusses the results of putting the above pieces together into MCMC simulations for parameter estimation of mock data corresponding to the burst model family of waveforms described below in Eq.~(\ref{eq:h_t}). 
From these simulations, in particular, we quantify the significant speed-ups that are obtained even for such a simple GW model when using the proposed ROQ. 
 
%%%%%%%%%%%%%%%%%%%%%%%%%%%%%%%%%%%%%%%%%%%%%
\section{Reduced Order Modeling}\label{sec:ROM}
%%%%%%%%%%%%%%%%%%%%%%%%%%%%%%%%%%%%%%%%%%%%%

Roughly speaking, ROM deals with data which can be represented by fewer degrees of freedom than those of the full problem with or without loss of accuracy. For a given problem there are many available methods for revealing a reduced representation. Classical methods such as Principal Component Analysis, Proper Orthogonal or Singular Value Decompositions (SVD) \cite{Pinnau2008}, which are related to each other, were introduced as early as the 1800's (see \cite{Stewart:1993:EHS:166597.166599} for a review of their history) and reveal low-rank approximations within existing data. Other approaches such as  Reduced Basis (RB) 
(see, as examples, Refs.~\cite{Maday:2002:PCT:608985.609022,Veroy2003619,prud'homme:70,FLD:FLD867,PateraReview,Nguyen_2009,Chen:2010:CRB:1958598.1958625,Knezevic20111455} or \cite{Quarteroni} for a recent review), are specifically designed for parametrized problems whose solution is expensive to evaluate but they also carry advantages when dealing with ``big data" problems (e.g. if the data cannot fit into memory or the SVD cost becomes prohibitive).

Both RB-greedy and SVD are projection-based ROM algorithms. If the waveforms are known at the training points
$$
{\cal T}:= \{\mlam_i \}_{i=1}^M
$$ 
with $\mlam_i$ some  parametrization of the samples, a projection-based method identifies a basis $\{ e_i\}_{i=1}^m$ such that 
\be
h(\cdot; \mlam ) \approx \sum_{i=1}^m c_i (\mlam ) e_i (\cdot ) \, , \quad \mbox{for } \mlam \in {\cal T} \label{eq:projapprox}
\ee
with $m\leq M$ and where the coefficients $c_i$ are given by Eq.(\ref{eq:optLS}) (see Appendix \ref{sec:RBapp} for more details). If the problem is amenable to ROM, then $m< M$ or even $m \ll M$.

To be more concrete, in the GW case $\mlam$ would represent the (intrinsic and/or extrinsic) parameters of the problem, and $M$ the number of available parameter samples; say, the number of waveforms in a catalog or even the continuum, $M \rightarrow \infty$. A generic waveform with associated parameter $\mlam$ would be a function of time or frequency, 
$$
h =  h(t; \mb{\lambda}) \quad \mbox{or} \quad h = h(f; \mb{\lambda}) \, . 
$$
In what follows, we will refer to $\mlam$ as the parameter dimension and $f$ or $t$ as the physical one.

%%%%%%%%%%%%%%%%%%%%%%%%%%%%%%
\subsection{Generating a basis}
%%%%%%%%%%%%%%%%%%%%%%%%%%%%%%

Suppose for any $\mb{\lambda}$ the GW template $h(\cdot ;\mb{\lambda})$ has an accurate approximation of the form (\ref{eq:projapprox}) in some basis $\{ e_i\}_{i=1}^m$. Recent work~\cite{Field:2011mf,Caudill:2011kv,PhysRevD.86.084046, Cannon:2011xk} has shown that for fixed but arbitrary physical and parameter ranges, a small number of basis functions is sufficient to accurately represent any waveform of the same physical model in that range. Furthermore, when the basis is generated through a RB-greedy algorithm (described in Appendix \ref{sec:RBapp}), the approximation error is guaranteed to yield a nearly optimal solution of the so-called n-width approximation problem~\cite{Binev10convergencerates,DeVore2012}. In the cases of interest this means exponential convergence of the representation error defined below in Eq.~(\ref{eq:greedyErr}) with respect to the number of basis functions, resulting in a very compact basis. In addition, the number of basis elements often exhibits negligible increase as the dimensionality of the problem grows \cite{PhysRevD.86.084046}.

Of the basis set $\{ e_i(\cdot ) \}_{i=1}^m$ we require $m$ to be small and the approximation to satisfy
\begin{align} \label{eq:greedyErr}
\sigma_m := \max_{\mlam } \min_{c_i \in \mathbb{C}} \left \| h(\cdot ;\mb{\lambda}) - \sum_{i=1}^m c_i(\mb{\lambda}) e_i(\cdot) \right\|^2 \leq \epsilon \, ,
\end{align}
where $\epsilon$ is a user defined bound for the error (in our cases, typically $\sim 10^{-12}$, see for example Fig.~\ref{fig:RBburst}), the coefficients $\{ c_i \}$ are chosen so as to optimize the approximant (see Appendix \ref{sec:RBapp}), and  the largest error in the parameter region of interest is taken. That is, $\sigma_m$ quantifies the error of the ``worst best'' approximation by the basis. 

Many possible basis choices, including traditional ones such as Chebyshev polynomials or Fourier basis, could satisfy the above required criteria. In practice, application-specific bases usually provide better accuracy for a given $m$ and also lead to a well-conditioned global interpolation procedure, as described in Sec.~\ref{sec:EIM}. 

We have mentioned the RB-greedy algorithm as one approach to generate a good basis. For definiteness, in the simulations of this paper our basis is constructed with such an algorithm (described in Appendix \ref{sec:RBapp}). Our proposed ROQ rule is, however, directly applicable to any projection-based ROM basis, including SVD~\cite{Cannon:poster,Cannon:2010qh,Cannon:2011xk}.

%%%%%%%%%%%%%%%%%%%%%%%%%
\subsection{An example of RB: burst waveforms} \label{sec:RBburst}
%%%%%%%%%%%%%%%%%%%%%%%%%

In order to illustrate our approach, we consider a  four parameter GW-burst waveform given by  the following sine-Gaussian waveform:
\begin{eqnarray}
h(t; \mlam ) := A e^{-(t-t_c)^2/(2\alpha^2)}\sin(2\pi f_0 (t-t_c)) \, ,
\label{eq:h_t}
\end{eqnarray}
where $A$, $f_0$ and $\alpha$ are the amplitude, frequency and width of the waveform respectively, and where  $t_c$ is the arrival time of the GW-burst signal and $t\in[-\infty,\infty]$. The Fourier transform (FT) of this waveform is given by
\begin{align}
{\tilde h} (f,t_c; \mlam )= e^{i2\pi f t_c}{\tilde h} (f; \mlam )\,,\label{h_f1}
\end{align}
where ${\tilde h} (f; \mlam )$ is the FT of the GW-burst at $t_c = 0$: 
 \begin{align}
{\tilde h} (f; \mlam ) = i2A\alpha \sqrt{2\pi} \sinh(4\pi^2\alpha^2f_0f) e^{-2\pi^2\alpha^2(f_0^2+f^2)} .
\label{h_f2}
\end{align}
This waveform family is described by four free parameters $\mb{\lambda} = (\alpha,f_0, t_c, A)$. We will build the RBs over just two parameters $(\alpha,f_0)$, since the others are extrinsic and can be handled differently, as discussed in Sec.~\ref{sec:ext_par}. 

We build the RB for these burst waveforms over the parameter space defined by
\be 
\alpha = [.02,2]\sec \quad \, , \quad f_0 = [.01,1]{\rm Hz} \, , \label{eq:range_pars}
\ee
sampled with 180 equally spaced training points in each dimension. Unless otherwise stated, the range given in Eq.~(\ref{eq:range_pars}) will be the default one for all experiments and the units will always be in seconds and Hertz. To represent any burst waveform drawn from  the above range we take
\be 
T = 32 \sec \quad , \quad f_\mathrm{s} = 64 {\rm Hz} \, , \label{eq:data_range}
\ee
to be our default observation time and sampling rate. Similarly, for the injected signals our default parameters will be  
\be
\alpha=1 \quad , \quad f_0=0.25 \quad , \quad t_c=0.1 \label{eq:injected} \, .
\ee
We will also present results for a two parameter model in which $t_c$ is fixed at $t_c=0$ and where $A$ is chosen to give a specified signal-to-noise ratio (SNR), $\rho$, with $\rho^2 = \langle h | h \rangle$ for the inner product defined by Eq.~(\ref{eq:scalarprod}).

\begin{figure}[ht]
\includegraphics[width=0.98\linewidth]{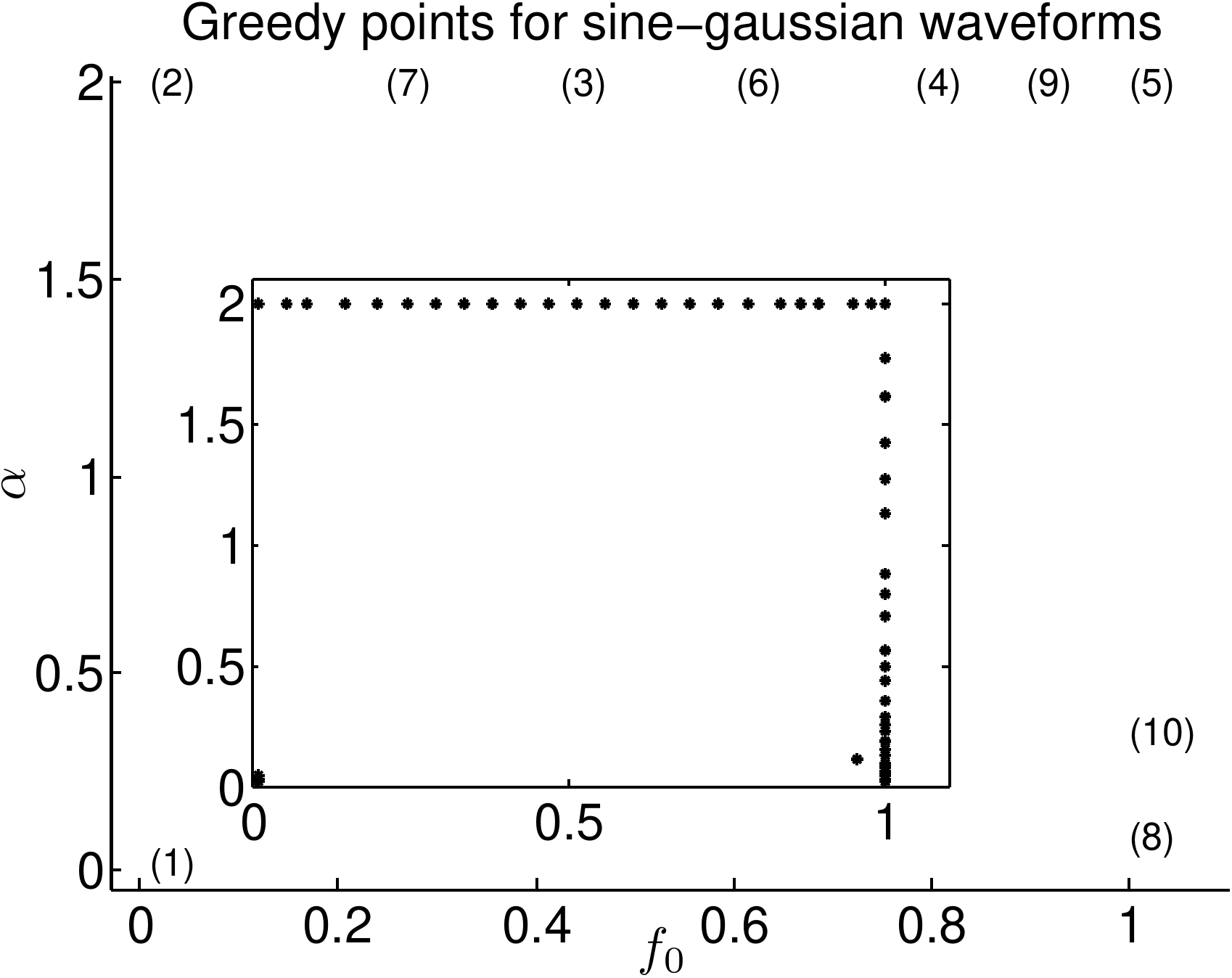}
\caption{Points selected by the greedy algorithm for the model family of burst waveforms (\ref{eq:h_t}) with the default range (\ref{eq:range_pars}) for its parameters. The first $10$ greedy points are represented with markers indicating the order of selection, with parenthesis serving as a visual aid. The inset figure shows with black asterisks all the $54$ selections, out of $180\times180$ samples, chosen by the greedy algorithm.}
\label{fig:greedypoints}
\end{figure}

\begin{figure}[ht]
\includegraphics[width=0.98\linewidth]{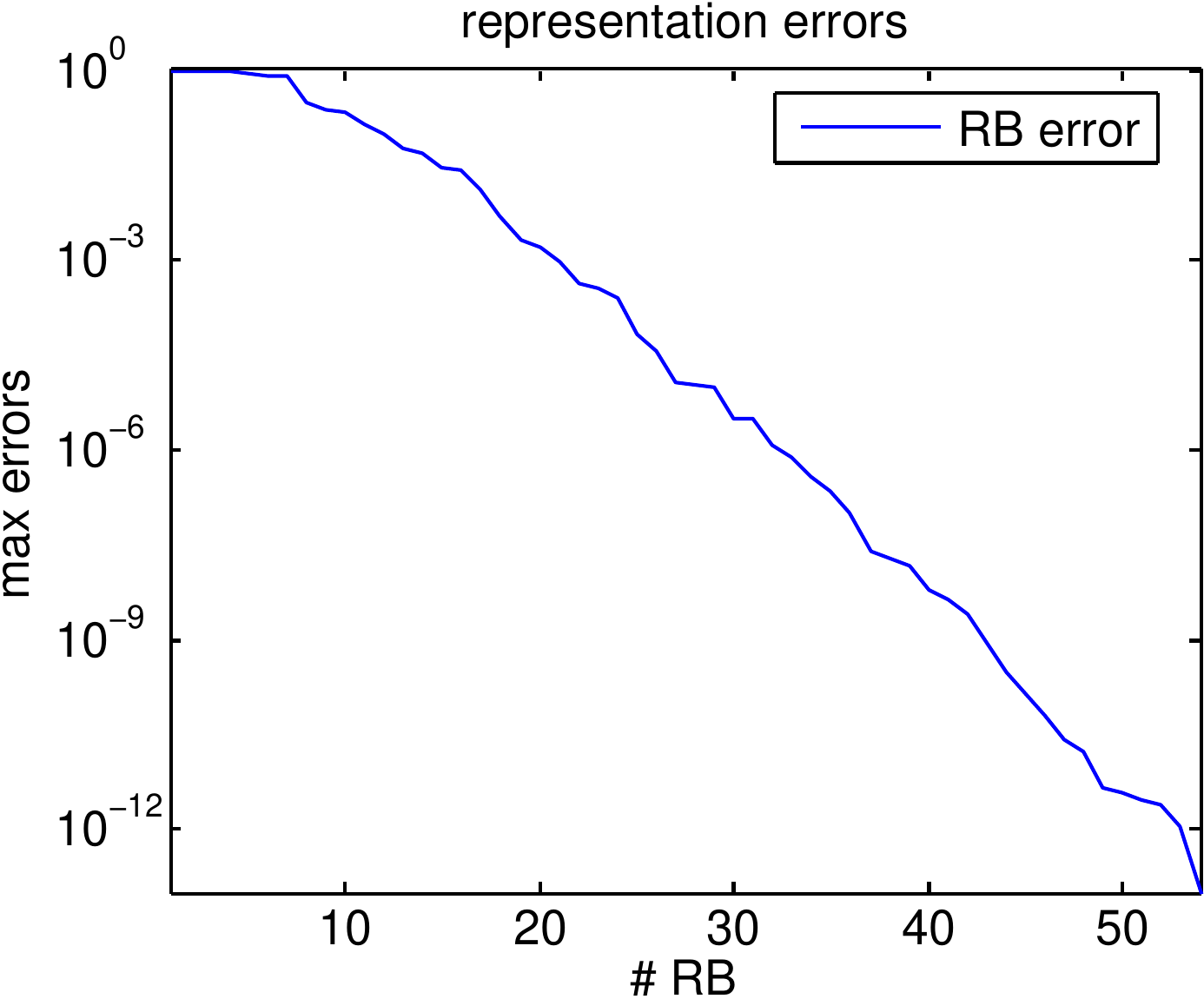}
\caption{Approximation error as a function of the number of basis generated with a greedy algorithm from the previous figure. 
The error $\sigma_m$, defined by Eq.~(\ref{eq:greedyErr}), is computed as the maximum 
within the parameter region given in Eq.~(\ref{eq:range_pars}).} 
\label{fig:RBburst}
\end{figure}

Fig.~\ref{fig:greedypoints} shows the $54$ points, out of $180\times 180$ samples, selected by the greedy algorithm to build the RBs, and the order in which  the first $10$ points are picked, while Fig.~\ref{fig:RBburst} shows the representation error of the training set as a function of the number of RB elements. Consistent with previous experience, we have found that if the training set is dense enough (and for this model, one of $180\times 180$ samples is) then any waveform not present in the training set yields similarly small representation errors by the basis; see for example \cite{Caudill:2011kv,PhysRevD.86.084046} for more details. 

So far we have described the generation of  basis elements. The next step is the {\em prediction} (as opposed to projection) of waveforms from a sparse set of well chosen frequency samples. 

%%%%%%%%%%%%%%%%%%%%%%%%%%%%%%%%%%%%%%%%%%%%%
\section{Empirical Interpolation} \label{sec:EIM}
%%%%%%%%%%%%%%%%%%%%%%%%%%%%%%%%%%%%%%%%%%%%%

Within a projection-based approximation one has
\be
h(x) \approx \sum_{i=1}^m c_i e_i (x) \label{eq:proj}\, , 
\ee
where the coefficients $c_i$ are given by Eq.~(\ref{eq:optLS}). Computing the projection coefficients $c_i$ requires full knowledge of the function $h$ (see Appendix \ref{sec:RBapp} for more details). 

Given a basis and partial sampling of $h$, in the interpolation problem we are interested in predicting the underlying function. In what follows, we will first review the classical interpolation problem, using a polynomial basis before discussing empirical interpolation with application-specific basis functions, and finish this section with an example for burst GWs.

\subsection{Classical interpolation with polynomials} \label{sec:interpolation}

Classically the interpolation problem for a function $h(x)$ is the following. Given a set of $m$ nodes $\{ x_{i} \}$, known function evaluations $\{ h_i:= h(x_i) \}$, and a basis $e_i = p_i(x)$ where $p_i(x)$ is a degree $i\leq m-1$ polynomial,
find an approximation (the interpolant) 
\be
{\cal I}_m[h](x) = \sum_{i=1}^m c_i p_i(x) \approx h(x) \label{eq:interp}
\ee
such that 
\be
{\cal I}_m[h](x_i) = h_i \quad \mbox{for} \quad i=1, \ldots, m\, .  \label{eq:interpb}
\ee
That is, the approximant is required to agree with the function at the set of $m$ nodes.

We can show that the problem defined by Eqs.~(\ref{eq:interp},\ref{eq:interpb}) has a unique solution in terms of Lagrange polynomials.
Given a convergence rate for the projection-based approximation Eq.~(\ref{eq:proj}) we might wonder how much accuracy is lost by trading it for the interpolation Eq.~(\ref{eq:interp}) and how to optimally choose the node points $x_i$. When the relevant error measurement is the maximum pointwise error, Chebyshev nodes are known to be well suited for interpolation, bringing an additional error which grows like $\log(m)$~\cite{Press92,Quarteroni2010}. 

For application-specific bases, a good set of interpolation points is not known a-priori. Next we describe an approach for identifying a nearly-optimal set.

\subsection{Empirical interpolation with RB}

The Empirical Interpolation Method was proposed in 2004 \cite{Barrault2004667} as a way of identifying a good set of interpolation points for arbitrary basis sets on multi-dimensional unstructured meshes and has since found numerous applications \cite{Maday_2009,chaturantabut:2737,Chaturantabut5400045,Eftang:2011,Aanonsen2009}. Recently, the EIM was shown to dramatically speed up parameterized inner product (overlap) computations in the absence of noise~\cite{antil2012two}. For definiteness we will focus on the frequency-domain case. In general, a well-posed interpolation problem for $m$ basis functions requires $m$ interpolation points $\{ F_i \}_{i=1}^m$. Additionally, these points must ensure an accurate approximation. Crucially, the EIM algorithm selects the interpolation points as a subset of the full $N/2+1$ data samples (this choice is motivated in Sec.~\ref{sec:Riemann}), $\{ F_i \}_{i=1}^m \subset \{ f_i\}_{i=0}^{N/2}$, and $m < N/2$ or even $m \ll N/2$.

With ROM we seek to find an empirical (that is, problem-dependent) {\em global} interpolant 
\be
{\cal I}_m [h](f;\mb{\lambda}) := \sum_{i=1}^m c_i (\mlam) e_i(f)\, ,  \label{eq:interpdef}
\ee
where the $c_i$ coefficients are defined as solutions to the interpolation problem
\begin{align} \label{eq:intproblem}
{\cal I}_m [h](F_k;\mb{\lambda}) = h(F_k;\mb{\lambda}) , \qquad \forall \, k=1,\dots,m .
\end{align}

For the moment, we shall assume that the EIM points are known (the precise way of finding them is explained in Appendix \ref{sec:EIMapp}) and proceed to describe how we use them to find the EIM interpolant.
Equation (\ref{eq:intproblem}) is equivalent to solving an $m$-by-$m$ system $A\vec{c} = \vec{h}$ for the coefficients $\vec{c}$, where
\begin{equation} \label{eq:InterpMatrix}
  A := \left(  \begin{array}{cccc}   
              e_1(F_1)  &  e_2(F_1)            & \cdots & e_{m}(F_1)      \\
              e_1(F_2)  &  e_2(F_2)            & \cdots & e_{m}(F_2)       \\
              e_1(F_3)  &  e_2(F_3)          & \cdots & e_{m}(F_3)   \\              
              \vdots    & \vdots             & \ddots & \vdots                       \\
              e_1(F_{m})  & e_2(F_{m})    & \cdots & e_{m}(F_{m})  \\               
             \end{array}
   \right) \, .
\end{equation}
The EIM algorithm ensures that the matrix $A$ is invertible, with $\vec{c} = A^{-1}\vec{h}$ the unique solution to Eq.~(\ref{eq:intproblem}). As $A$ is parameter independent we have, for all values of $\mb{\lambda}$, 
\begin{align} \label{eq:DEIM}
{\cal I}_m [h](f;\mb{\lambda}) = \vec{e}^{\hspace{2pt}T}(f) \left[ A^{-1}\vec{h}(\mb{\lambda}) \right] ,
\end{align}
where $\vec{e}^{\hspace{2pt}T} = [e_1(f),\dots,e_m(f)]$ denotes the transpose of the basis vectors, which we continue to view as functions.  

The empirical interpolant is nearly optimal in the sense that it satisfies
\begin{align}
\max_{\mlam } \| h(\cdot ;\mb{\lambda}) - {\cal I}_m [h(\cdot ;\mb{\lambda})] \|^2 \leq \Lambda_m^2 \sigma_m \, , \label{eq:errorEIM}
\end{align}
where $\sigma_m$ characterizes the representation error of the basis as defined in Eq.~\eqref{eq:greedyErr} and $\Lambda_m$ is a computable Lebesgue constant. For more details and in the context of GWs, see, for example, \cite{antil2012two}. For problems with smooth dependence with respect to parameter variation we can expect exponential decay of $\sigma_m$ with respect to $m$ and therefore of the EIM error (\ref{eq:errorEIM}) as well. 

%%%%%%%%%%%%%%%%%%%%%%%%%
\subsection{An example of EIM: burst waveforms}
%%%%%%%%%%%%%%%%%%%%%%%%%

We now provide a qualitative outline of the EIM algorithm, with more details given in Appendix \ref{sec:EIMapp}. As input the algorithm takes the basis set $\{ e_i \}_{i=1}^m$ and an arbitrary number and choice of data samples $\{ f_i\}_{i=0}^{N/2}$ from which the empirical interpolation points $\{ F_i \}_{i=1}^m$ are to be selected. The EIM algorithm proceeds as follows
\begin{enumerate}
\item The first point is chosen to maximize the value of $| e_1(f_i) |$; that is, $\left| e_1(F_1) \right| \geq \left| e_1(f_i) \right|$ for all data samples. 
\item Next,  an empirical interpolant for the second basis function is built using only the first basis function: From Eqs.~(\ref{eq:interpdef},\ref{eq:intproblem}) or, equivalently, Eq.~\eqref{eq:DEIM} we have ${\cal I}_1 [e_2](f)= c_1 e_1(f)$ where $c_1 = e_2(F_1) / e_1(F_1)$ has been found from Eq.~\eqref{eq:intproblem} with $k=1$. 
\item The second empirical interpolation point is chosen to maximize the value of the pointwise interpolation error of ${\cal I}_1 [e_2](f) - e_2(f)$; that is,  
$\left| {\cal I}_1 [e_2] (F_2) - e_2(F_2) \right| \geq \left| {\cal I}_1 [e_2](f_i) - e_2(f_i) \right|$ for all data samples. 
\item Steps $2$ and $3$ are then repeated to select the remaining $m-2$ points. 
\end{enumerate}

As described, the EIM follows a greedy approach, albeit somewhat different from that one we used to build a Reduced Basis. While a greedy algorithm to build a RB selects the most relevant points in parameter space, the EIM selects the most relevant points in the {\em physical dimension(s)}. 

\begin{figure}[ht]
\includegraphics[width=0.78\linewidth]{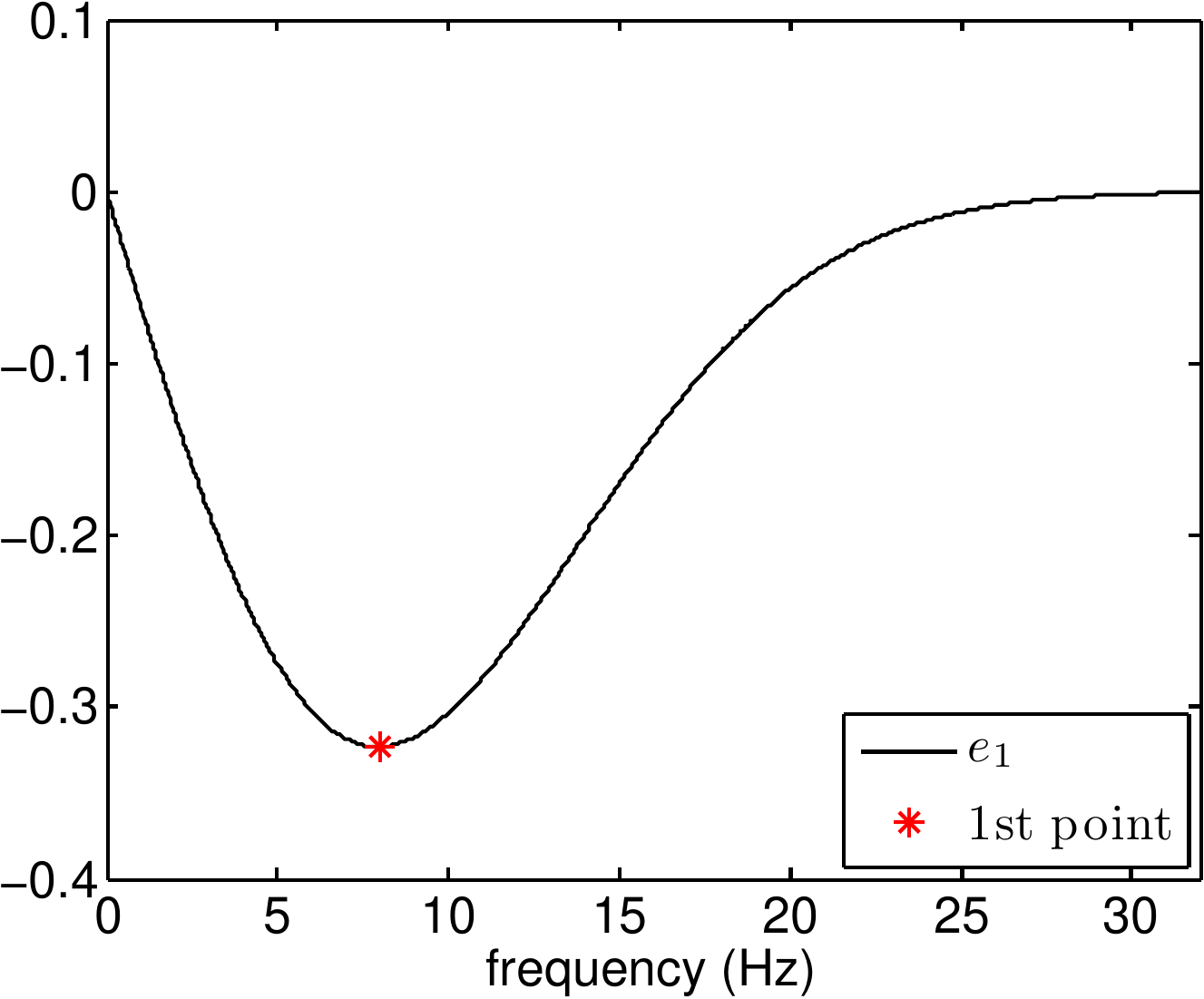} \\
\includegraphics[width=0.78\linewidth]{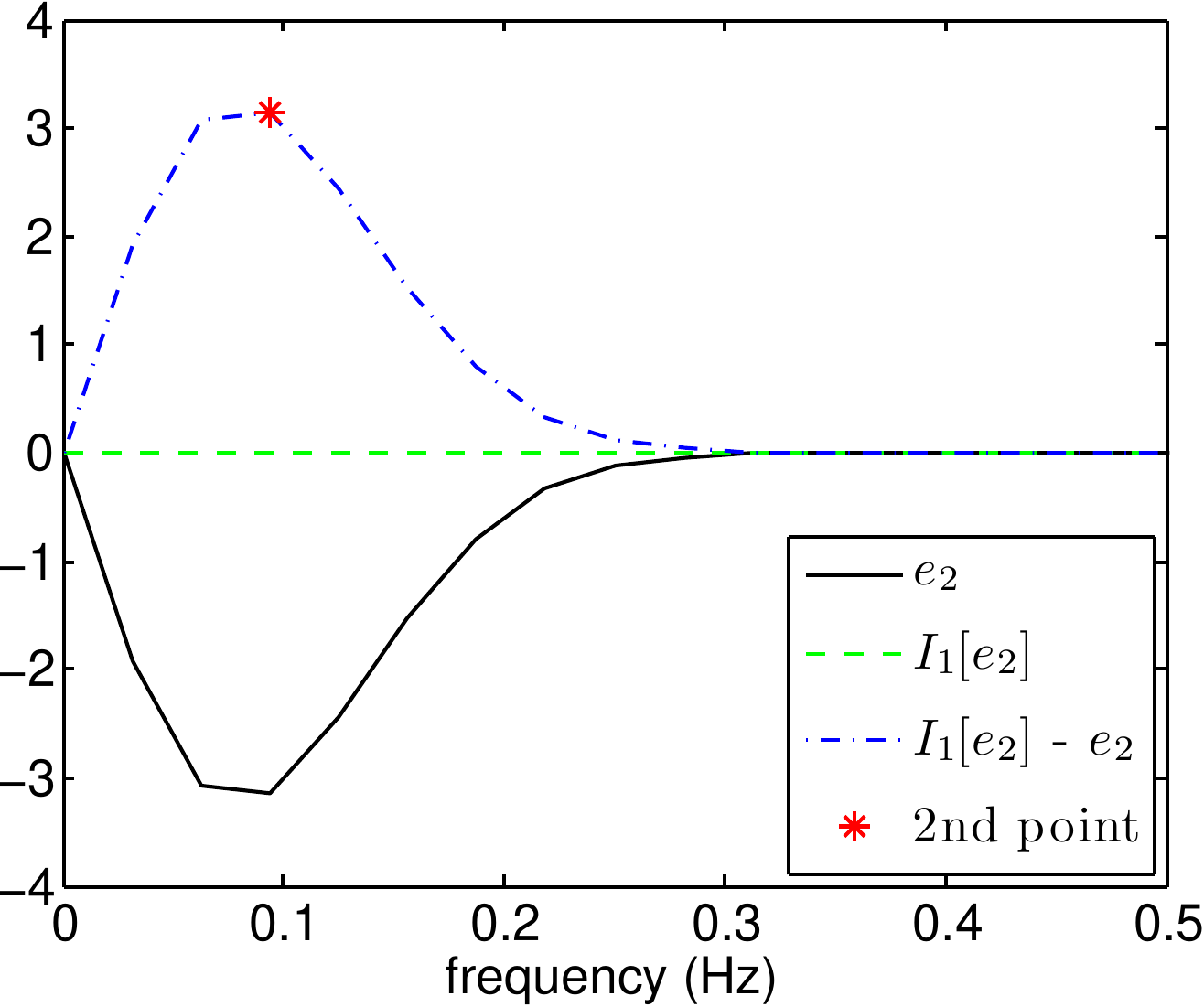}
\caption{Iterations 1 (top) and 2 (bottom) of the EIM algorithm. The first EIM point is defined by the location of max$(|e_1|)$. To identify the second point we: i) build the empirical interpolant ${\cal I}_1[e_2]$ of $e_2$ using $e_1$ and the sample point $F_1$ (cf Eq.~\eqref{eq:DEIM}), ii) compute the pointwise error ${\cal I}_1[e_2] - e_2$; iii) the second EIM point is then defined by the location of max$(\left|{\cal I}_1[e_2] - e_2\right|)$. The process continues until all $m$ empirical interpolation points are found.
\label{fig:deimalg}}
\end{figure}

\begin{figure}[ht]
\includegraphics[width=0.98\linewidth]{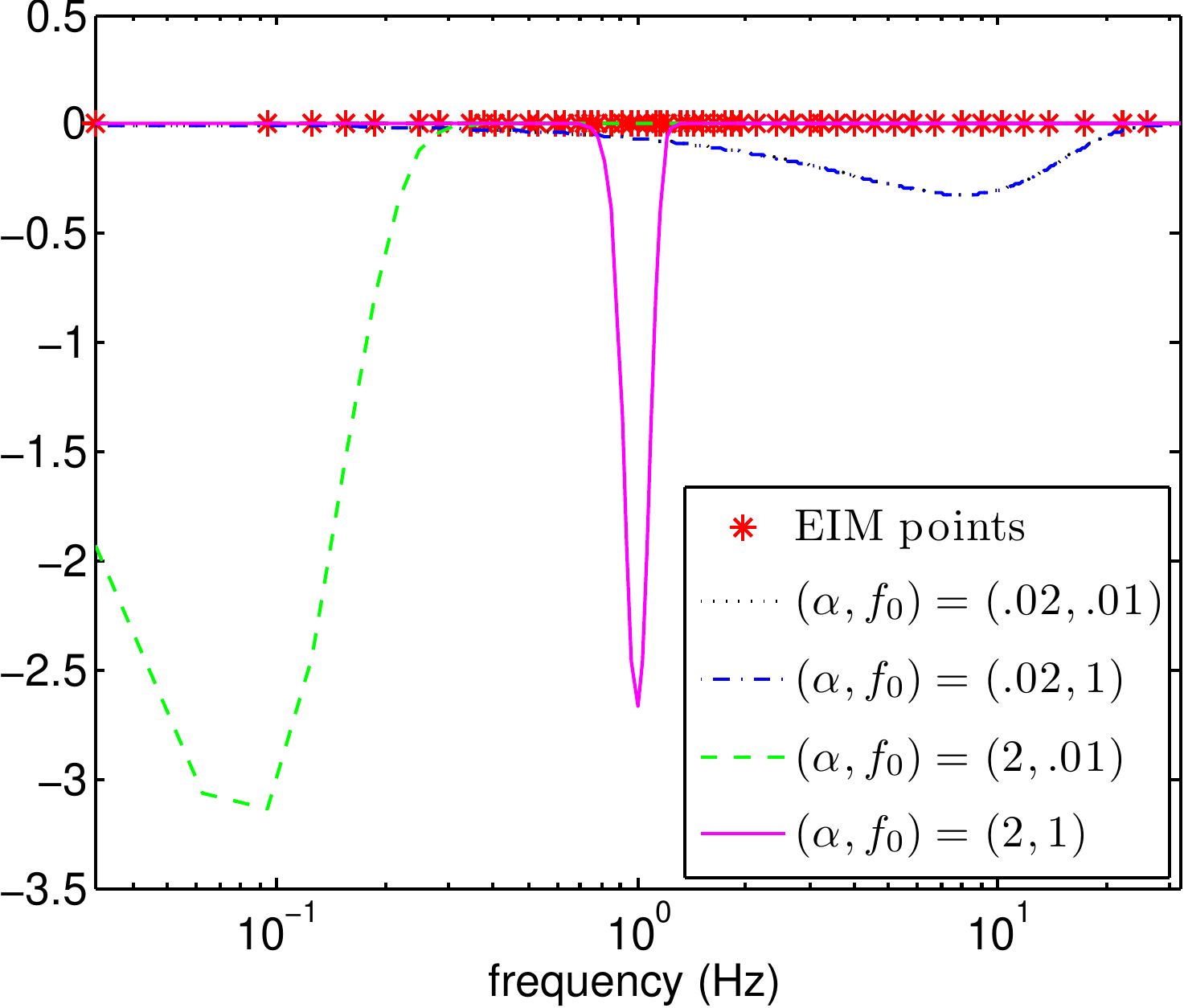}
\caption{Empirical interpolation points (red asterisks) selected by the EIM algorithm for the sine-Gaussian waveforms. These points are a subset of the original data (which in this case has equidistant spacing $\Delta f$, see Sec.~\ref{sec:Riemann}) and cluster towards lower $f \sim 1$Hz, as expected. Four representative waveforms are depicted for all possible combinations of max/min values of the waveform frequency $f_0$ and width $\alpha$. Greater diversity in waveform features is evident at lower frequencies. 
\label{fig:deimpoints}}
\end{figure}

\begin{figure}[ht]
\includegraphics[width=0.98\linewidth]{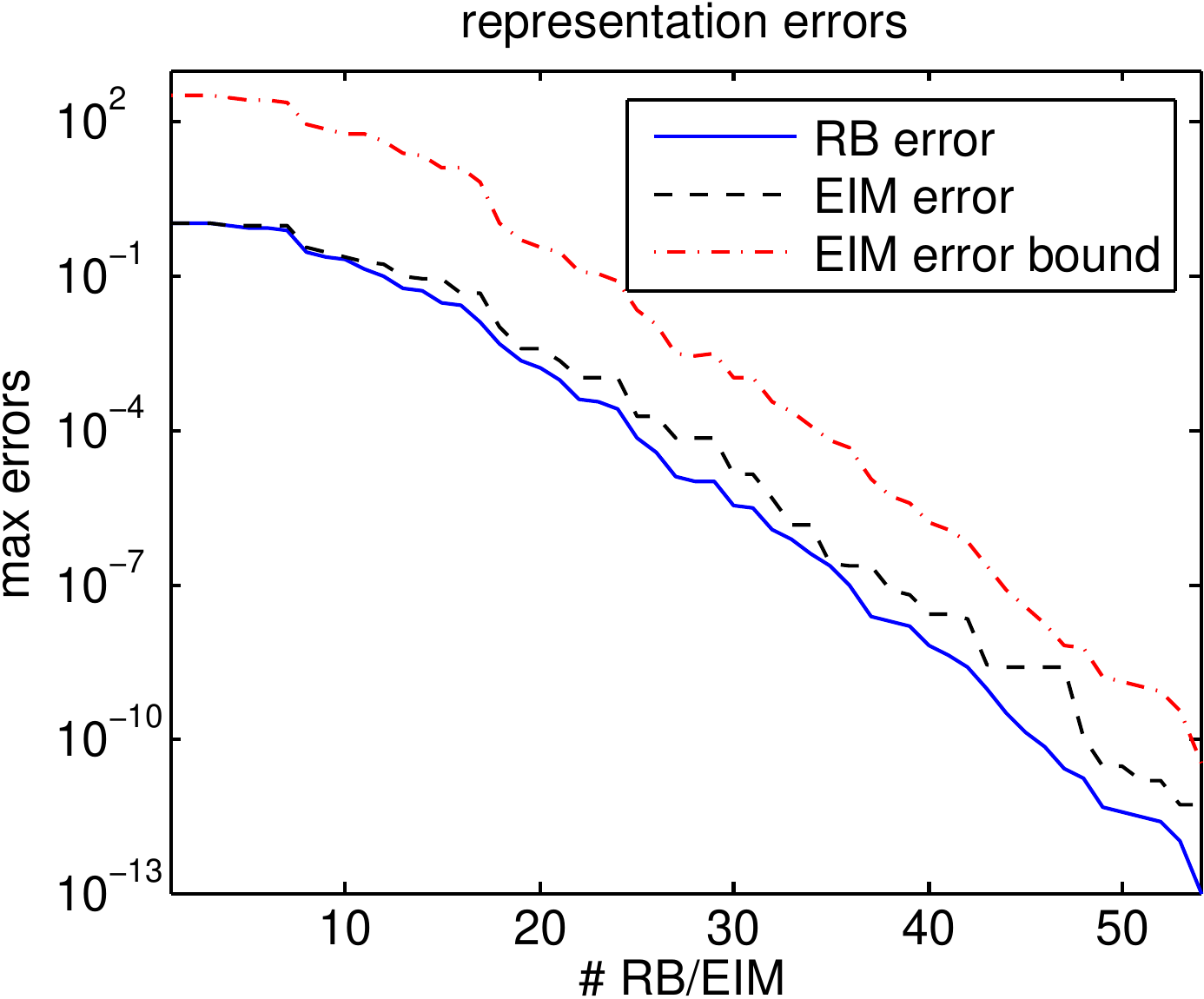}
\caption{Approximation error as a function of the number of Reduced Basis (RB) generated with a greedy algorithm (solid blue), and for the Empirical Interpolant (dashed black), defined as $\sigma_m$ and $\max_{\mlam} \|h - {\cal I}_m[h] \|^2$, respectively. The dashed red line shows the error bound [see Eq.~(\ref{eq:errorEIM})].  }
\label{fig:reperrs}
\end{figure}

Fig.~\ref{fig:deimalg} provides a graphical illustration of the EIM algorithm's first iterations for the family of sine-Gaussian burst waveforms (\ref{h_f2}), using the RB described in Sec.~\ref{sec:RBburst}. All $m=54$ point selected by the greedy algorithm (see Sec.~\ref{sec:RBburst}) are shown in Fig.~\ref{fig:deimpoints}.  Finally, in Fig.~\ref{fig:reperrs} we show the largest empirical interpolation error of $10,000$ waveforms drawn randomly from the parameter region [Eq.~\eqref{eq:range_pars}].
 
%%%%%%%%%%%%%%%%%%%%%%%%%%%%%%%%%%%%%%%%%%%%%
\section{Reduced Order Quadratures} \label{sec:ROQ}
%%%%%%%%%%%%%%%%%%%%%%%%%%%%%%%%%%%%%%%%%%%%%
As anticipated and summarized in Sec.~\ref{sec:overview},  building an ROQ has {\it offline} and {\it startup} costs, with the advantage of very fast {\it online} evaluations. In the offline stage we construct the basis and EIM points. This stage is independent of any data/signal. The startup stage, in turn, is data-dependent and completes the ROQ, which preserves the accuracy of any quadrature rule of interest with a number of quadrature nodes which equals the number of basis functions. Roughly speaking, the accuracy of the resulting ROQ is comparable to that of the basis, with the nodes chosen as a subset of the data points at which the signal has been sampled.

The details of how to construct an ROQ rule mimic well known quadratures rules. Let us briefly recall how these standard quadratures are derived for the integration of a real function $h(x)$: the function is approximated by its polynomial interpolant (cf.~Sec.~\ref{sec:EIM}) and the latter integrated exactly to compute the weights of the rule. Namely, given the interpolation approximation 
$$
h(x) \approx \sum_{i=1}^m h(x_i) \ell_i(x) \, , 
$$
where $\ell_i(x)$ are Lagrange polynomials (see Sec.~\ref{sec:interpolation}), standard quadratures are derived as 
$$
\int h(x) dx \approx \sum_{i=1}^m h(x_i) \alpha_i  \qquad \alpha_i := \int \ell_i(x) \, .
$$
Interpolation at equally spaced points for $m=1$ leads to the trapezoidal rule, for $m=2$ to Simpson's rule, etc. By additionally choosing the location of the interpolation points we can maximize the exactness of the quadrature rule for polynomials, leading to Gaussian quadratures. 

%%%%%%%%%%%%%%%%%%%%%%%%%%%%%%%%%%%%%%%%%%%
\subsection{Riemann sum with uniform sampling} \label{sec:Riemann}
%%%%%%%%%%%%%%%%%%%%%%%%%%%%%%%%%%%%%%%%%%%
In general, the output of a GW detector is comprised of data segments of duration $T$, which are uniformly sampled every $\Delta t$ seconds. Assuming for simplicity $t_c=0$ for the time being (how to include the arrival time is discussed in Sec.~\ref{sec:ext_par}), for $N = T/( \Delta t)$ data samples the discrete GW waveform 
$$
h(j\Delta t;\mb{\lambda}) \, , \quad j=0\ldots N\,,
$$
has discrete FT  $\tilde{h}(f_i;\mb{\lambda})$, which is known at the frequency points $\{ f_i\}_{i=0}^{N/2} = \{0, f_0 , 2f_0 , \dots , \left( N/2 \right) f_0\}$, where $f_0=1/T=\left( N \Delta t\right)^{-1} = \Delta f $ is the fundamental frequency and $f_\mathrm{max} = (N/2) f_0$. 

Due to the fact that the data taking procedure dictates the instants of time at which the (non-smooth and noisy) signal is known, 
an obvious numerical approximation to Eq.~\eqref{eq:scalarprod} is a low order discrete Riemann sum, 
\begin{align} \label{eq:NumRiem}
\left<a\left|b\right.\right> 
\approx \left<a\left|b\right.\right>_{\text{d}}  := \frac{4}{N \Delta} \Re \sum_{i=0}^{N/2} \left[ \frac{\tilde{a}(f_i) \tilde{b}^{*}(f_i)}{\tilde{S}_{n}(f_i)} \right]  \,  . 
\end{align}
Thus, the computational cost of Eq.~(\ref{eq:NumRiem}) depends on $N$, which in turn depends on the data sampling rate. 

Whether performing searches or parameter estimation studies, the numerical integral Eq.~\eqref{eq:NumRiem} is repeatedly evaluated for a variety of GW templates $h(f,\mb{\lambda})$. Next we show how such integrals can be computed with a cost not set by the Nyquist sampling rate, but rather the ``information content" of the GW templates themselves; namely, the number of basis functions, $m$. This is similar in spirit to the fact that compressed sensing can ``beat" Nyquist-Shannon sampling criteria~\cite{daCostaRibeiro:2011:CSS:2186225.2186298}. 
\newline

%%%%%%%%%%%%%%%%%%%%%%%%%%%%%%%%
\subsection{Building the ROQ}
%%%%%%%%%%%%%%%%%%%%%%%%%%%%%%%%

Consider a discrete approximation $\langle \cdot | \cdot \rangle_{\tt d}$ to the continuum scalar product of Eq.~(\ref{eq:scalarprod}). The Riemann sum Eq.~(\ref{eq:NumRiem}) is a natural choice in data analysis studies, whether for Bayesian parameter estimation or searches with matched filtering. Given the discrete FT of a data set $\tilde{s}(f_i)$ (one can similarly build an ROQ in the time domain), and specializing to white noise $\tilde{S}_n = 1$ without loss of generality (one can absorb $\tilde{S}_n$ into the definition of $\tilde{s}$), ROQ inner products between data and templates $h(f;\mb{\lambda})$ are computed as
\begin{widetext}
\begin{eqnarray} \label{eq:ROQ}
\langle h(\mlam ) | s  \rangle_{\text{d}} &= &  4 \Re \sum_{k=0}^{N/2} s^{*}(f_k) h(f_k;\mb{\lambda}) \Delta f \nonumber \\
& \approx & 4  \Re \sum_{k=0}^{N/2} s^{*}(f_k) \cI_m [h(f_k;\mb{\lambda})] \Delta f  = 
4 \Re \sum_{k=0}^{N/2} s^{*}(f_k) \left[ \vec{e}^{\hspace{2pt}T}(f_k) A^{-1}\vec{h}(\mb{\lambda}) \right] \Delta f \nonumber  \\
& = &  4 \Re \left[ \sum_{k=0}^{N/2} s^{*}(f_k) \vec{e}^{\hspace{2pt}T}(f_k)  \Delta f  A^{-1} \right]\vec{h}(\mb{\lambda}) 
=  4 \Re \sum_{k=1}^{m} \omega_k h(F_k;\mb{\lambda}) \nonumber \\ 
& =: & \langle h(\mlam ) | s  \rangle_{\text{\tiny{ROQ}}}  \,,  \nonumber
\end{eqnarray}
\end{widetext}
where the coefficients  $\omega_j$  are given by:
\be
\omega_j := \sum_{k=0}^{N/2} s^{*}(f_k) e_j(f_k)  \Delta f  A^{-1}  \, . \label{eq:roq_weights}
\ee
The vector 
$$
\sum_{k=0}^{N/2} s^{*}(f_k) \vec{e}^{\hspace{2pt}T}(f_k) \Delta f 
$$
is composed of inner products between all the basis elements and the data. We refer to $\{\omega_k\}_{k=1}^m$ Eq.~(\ref{eq:roq_weights}) as {\em data-specific weights}, and their generation comprises the ROQ {\it startup cost}. Defining the scalar product between the data and the $j^{th}$ basis function by 
\begin{align} \label{eq:innerwithbasis}
E_j := \sum_{k=0}^{N/2} s^{*}(f_k) e_j(f_k) \Delta f  \, , 
\end{align}
the data-specific weights are given by
\begin{align} \label{eq:weights}
\vec{\omega}^{\hspace{2pt}T} = \vec{E}^{\hspace{2pt}T} A^{-1} \, .
\end{align}
Notice that the ROQ nodes are exactly the EIM points which, together with the weights (\ref{eq:roq_weights}), completes our ROQ approximation  
\begin{align} \label{eq:ROQoverlap}
\langle h(\mb{\lambda}) | s \rangle_{\text{\tiny{ROQ}}} = 4 \Re \sum_{k=1}^{m} \omega_k h(F_k;\mb{\lambda})   \, .
\end{align}
The ROQ rule's accuracy only depends on the interpolant's accuracy to represent $h(f;\mb{\lambda})$ and the accuracy of the original quadrature $\langle \cdot | \cdot \rangle_{\tt d}$. In particular, the method does {\em not} assume $s$ to be well approximated by the basis (i.e. neither waveform  modeling assumptions nor details about the noise realization are important).  Since, as discussed, the error of the interpolant, Eq.(\ref{eq:errorEIM}) can be expected to decay exponentially for the cases of interest, in practice the ROQ replaces the original quadrature rule by a less expensive one with the same accuracy (within, say, machine precision). How much smaller $m$ is compared to $N/2$ is model-dependent; in Sec.~\ref{sec:MCMC_ROQ} we quantify this for the family of burst waveforms described in Eq.~(\ref{h_f2}).

\begin{figure}[ht]
\includegraphics[width=0.98\linewidth]{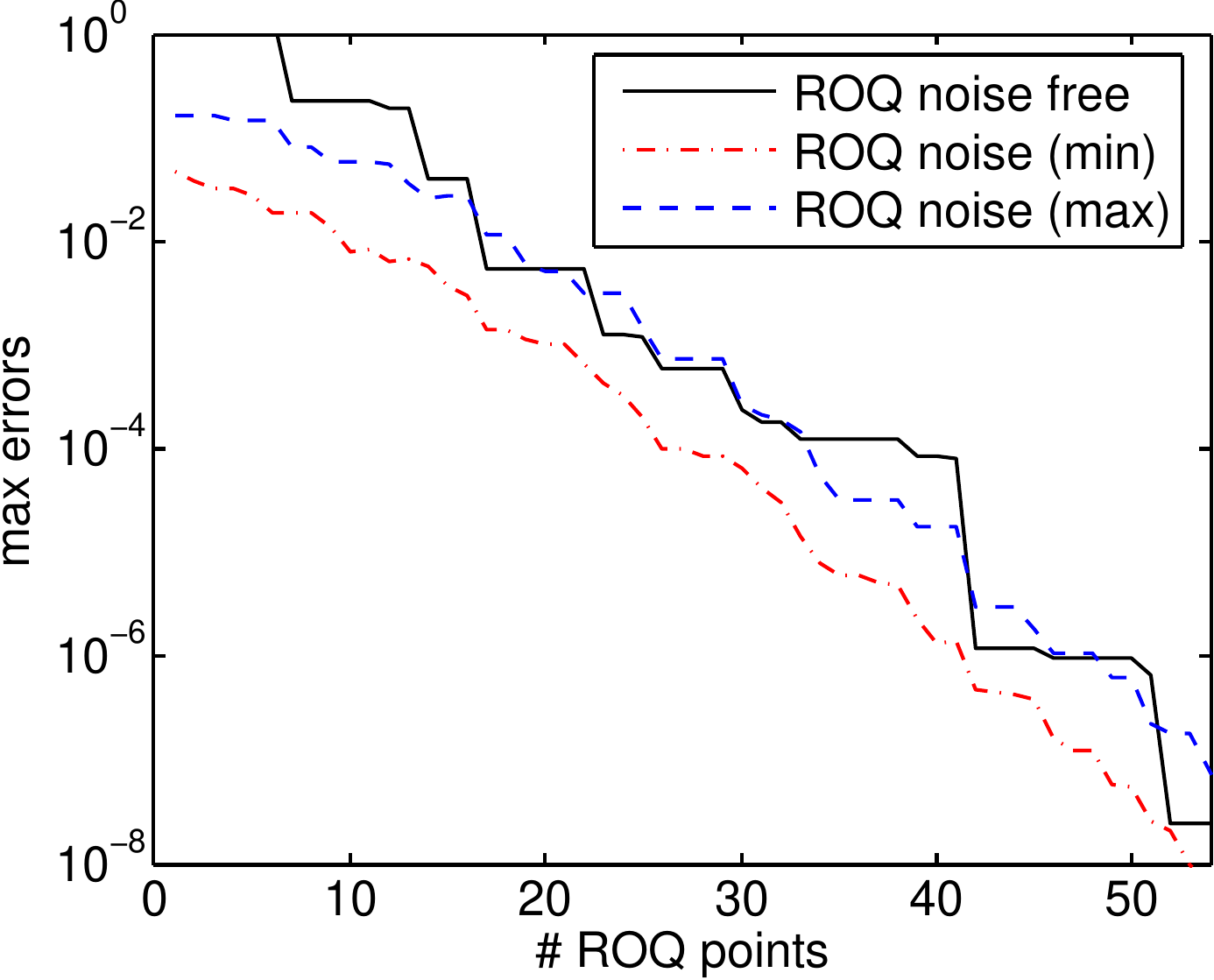}
\caption{Integration error (\ref{eq:int_err}) versus number of ROQ nodal points ($m$) for $10,000$ randomly selected values of $\mb{\lambda}$. The solid black curve depicts the noise-free case $s = h$ and the last data point $\left( m= 54, \sigma_m \approx 3 \times 10^{-8} \right)$ corresponds to the rule used in Fig.~\ref{fig:errorstc}. The blue and red curves show the maximum and minimum, over $100$ realizations of pure noise data, $s = n$, of the error maximised over all parameter values $\mlam$. {\em Note that the ROQ shows exponential convergence with respect to the number of ROQ nodes, even for pure noise data}.  
\label{fig:roqerr}}
\end{figure}

Figure~\ref{fig:weights} shows the nodes chosen by the EIM in the frequency domain, and the ROQ weights (\ref{eq:weights}) for the burst waveforms (\ref{h_f2}).  Figure~\ref{fig:roqerr}, in turn, shows the error
%-------
\begin{eqnarray}
\left| \langle h(\mb{\lambda}) | s \rangle_{\text{\tiny{ROQ}}} - \langle h(;\mb{\lambda}) | s \rangle_{\tt d} \right|\,,\label{eq:int_err}
\end{eqnarray}
%------
which arises in the computation of the overlap, in both cases with and without noise. Here the {\it max errors} label on the vertical axis refers to the maximum error found in a thorough sampling of the parameter range (\ref{eq:range_pars}), and the error is relative to a standard Riemann-sum integration (\ref{eq:NumRiem}) with $1025$ points.

\begin{widetext}
\begin{center}
\begin{figure}[ht]
\includegraphics[width=0.65\linewidth]{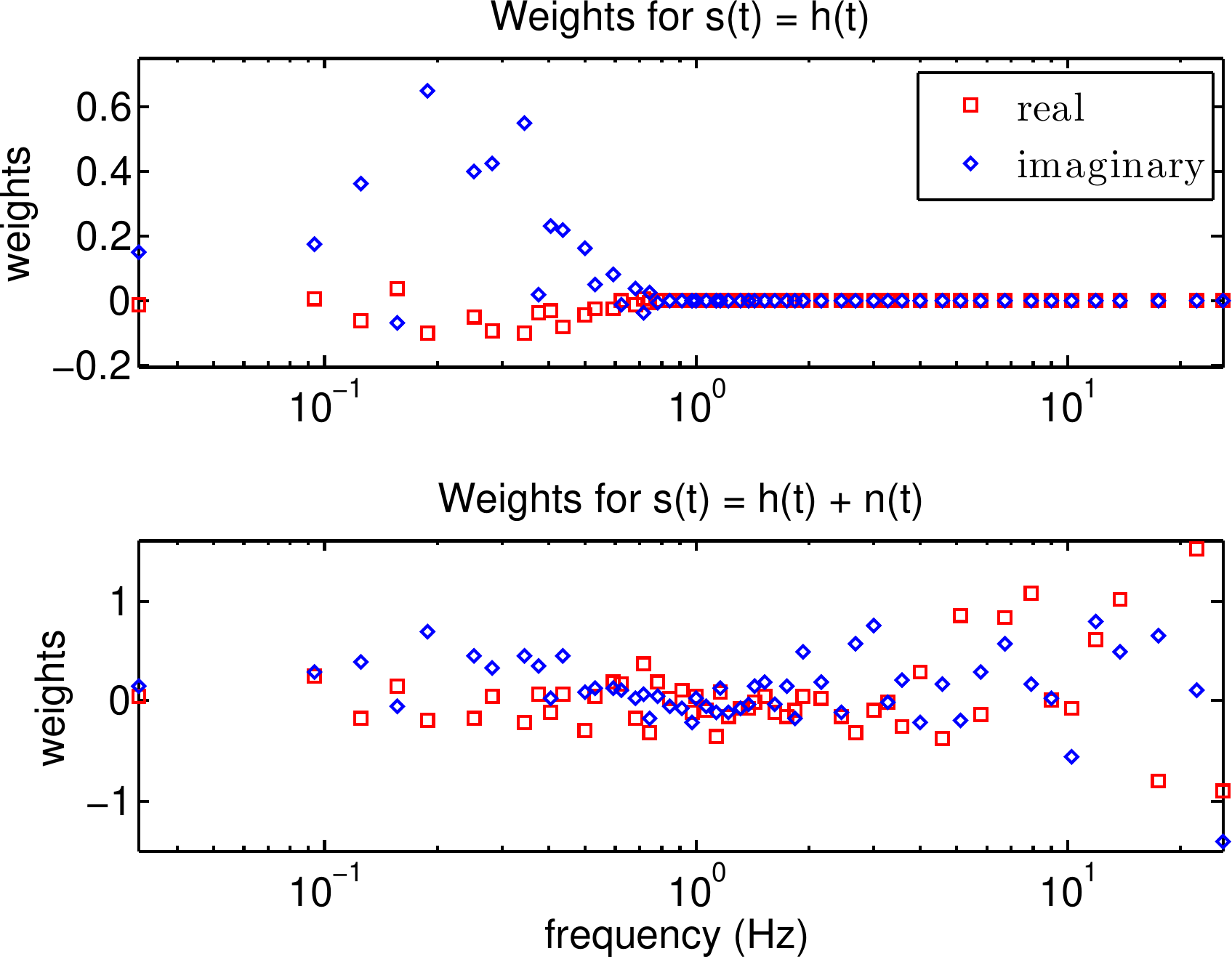}
\caption{Real (red squares) and imaginary (blue diamonds) ROQ weights computed from Eq.~(\ref{eq:weights}) for the test burst family waveforms in the range given by Eq.~(\ref{eq:range_pars}) and the injected signal with default parameters (\ref{eq:injected}). The top figure is for the noise-free case when $s(t) = h(t)$ while the bottom figure shows weights when $s(t) = h(t) + n(t)$, where $n(t)$ is a particular noise realization. 
\label{fig:weights}}
\end{figure}
\end{center}
\end{widetext}

%%%%%%%%%%%%%%%%%%%%%%%%%%%
\subsection{Extrinsic parameters}\label{sec:ext_par}
%%%%%%%%%%%%%%%%%%%%%%%%%%%
So far we have described how to build ROQs over the intrinsic parameters characterizing the waveform signal. The extrinsic parameters include the arrival time of the signal $t_c$ \footnote{Which, for the model (\ref{eq:h_t}), we take to be the time at the midpoint of the burst.}, the phase of the waveform at this time, and parameters such as the sky position, orientation and distance to the source. The phase of the waveform affects the model simply as multiplication by a complex constant, which keeps the waveform in the RB space. Similarly, sky position, orientation and distance just affect the amplitude of the source and the projection of the plus and cross polarizations of the waveform into a detector response and also do not take the waveform out of the RB space. However, the arrival time $t_c$ requires some more discussion. 

If we denote by $\mlam$ the set of parameters excluding $t_c$ and by $h_0(t; \mlam)$ the waveform computed with $t_c = 0$, then
\begin{equation}
h(t; t_c, \mlam) = h_0(t-t_c; \mlam)\,,
\end{equation}
with FT given by  ${\tilde h} (f;t_c, \mlam )$ [ see Eq.~(\ref{h_f1})].
For parameter estimation we compute integrals of the form \begin{equation}
O(t_c,{\bf\lambda}) := \int_0^\infty \frac{\tilde{h}(f; t_c, {\bf\lambda}) \tilde{s}^*(f)}{S_n(f)} \,\, {\rm d}f .
\end{equation}
The simple dependence of the FT is exploited in GW searches by defining the function $\tilde{I}_0(f; {\bf\lambda})$ via
\begin{equation}
\tilde{I}_0(f; {\bf\lambda}) = \frac{\tilde{h}_0(f; {\bf\lambda}) \tilde{s}^*(f)}{S_n(f)}
\end{equation}
for which 
\begin{equation}
O(t_c,{\bf\lambda}) = \int_0^\infty \tilde{I}_0(f; {\bf\lambda}) {\rm e}^{2 \pi{\rm i}ft_c} {\rm d} f = 2\pi I_0(-t_c; {\bf\lambda}) \, , 
\label{tcolap}
\end{equation}
where $I_0(t; {\bf\lambda})$ is the inverse FT of $\tilde{I}_0(f; {\bf\lambda})$. Since fast Fourier transforms are efficient, we can search over $t_c$ cheaply by doing this inverse FT.

The ROQ rule that we have computed for waveforms $h_0(t; {\bf\lambda})$ enables us to compute the integral of $\tilde{I}_0(f; {\bf\lambda})$ cheaply. However, we now need to compute the integral of $\tilde{I}_0(f; {\bf\lambda})\exp(2 \pi{\rm i}ft_c)$ and so the existing ROQ rule is in principle not guaranteed to work. However, if the ROQ is being used for follow-up parameter estimation, this will normally be triggered by the detection of a candidate event in the data stream of one or more detectors. These triggers will normally be able to localize the event to within a time interval comparable to a couple of cycles of the signal.

\begin{figure}[ht]
\includegraphics[width=0.98\linewidth]{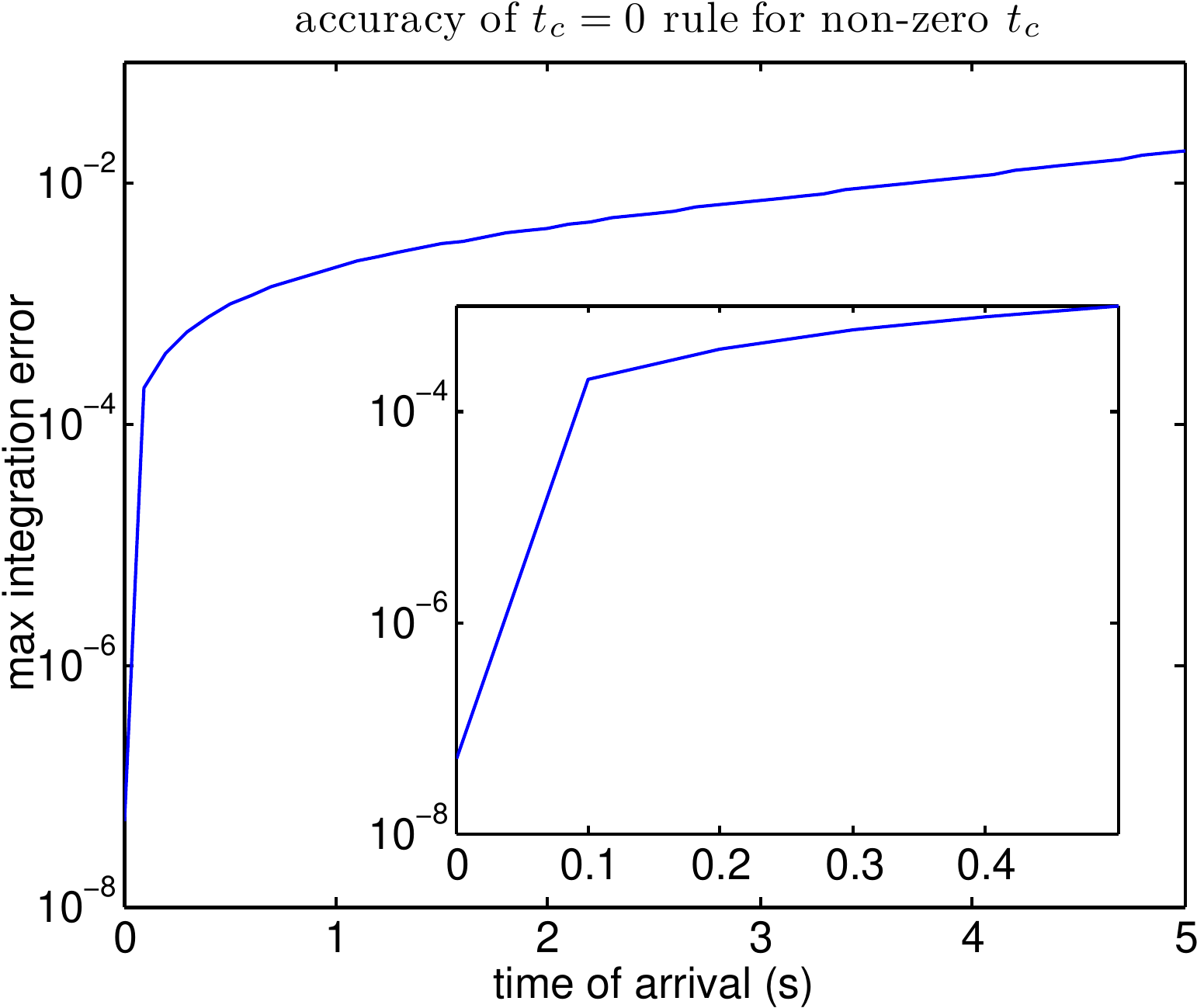}
\caption{Errors in computing the correlation between the data stream $s$ and the model waveform $h$ (see Sec.\ref{sec:overview}) $\left<s\left|h(\mb{\lambda},t_c)\right.\right>$  using an ROQ rule built for $t_c = 0$ with accuracy better than $\sim 10^{-6}$. Empirically we find that this rule continues to work well for non-zero values of $t_c$. Looking ahead to Sec.~\ref{sec:MCMC_ROQ} we anticipate evaluating the likelihood function for $t_c \leq 0.5 \sec$.
\label{fig:errorstc}}
\end{figure}

In practice, the simplest approach to handling $t_c$ is to build an ROQ rule for an estimated value (which we can denote by $t_c=0$ without loss of generality) and use it for other arrival times within a reasonable window around that value. In this way we can include the arrival time information {\em at no extra cost}. We show the error that arises from using a ROQ built for $t_c=0$ for non-zero values of $t_c$ in Fig.~\ref{fig:errorstc}. If higher accuracy is desired, we can build an ROQ which includes $t_c$ within the parameter space without losing efficiency, since it is an offline computation. Due to the fact that an estimate of  the  prior of the $t_c$ is known, and typically small, we have found that, as it was expected, the number of basis (and therefore ROQ nodes) increases by a small amount. Alternatively, we can build ROQ weights $\vec{\omega}\left( t_c \right)$  for different values of $t_c$ from Eq.~(\ref{eq:roq_weights}), increasing the startup cost, and interpolate $\omega_i(t_c)$ in $t_c$. We have found that these coefficients have a weak dependence on $t_c$ making them simple to interpolate.

\subsection{Computing the likelihood}

In order to evaluate the likelihood we compute Eq.~\eqref{eq:chi} as
\begin{align}
\langle s | s \rangle + \langle h\left(\mb{\lambda}\right) | h\left(\mb{\lambda}\right) \rangle - 2 \Re \langle s | h\left(\mb{\lambda}\right) \rangle \, ,
\end{align}
where the last term is handled with the ROQ rule \eqref{eq:ROQoverlap} and the first term needs to be computed once. In the case that the data stream $s(t)$ contains a sine-Gaussian burst-waveform (\ref{eq:h_t}) and  white noise $n(t)$ (see Sec.\ref{sec:overview}), we can compute a closed-form expression for the norm, 
\begin{equation}
\langle h\left(\mb{\lambda}\right) | h\left(\mb{\lambda}\right) \rangle = 4 A^2 \alpha\sqrt{\pi} \left( 1-e^{-4\pi^2f_0^2\alpha^2}\right)\, , 
\end{equation}
where $f_\mathrm{min} = 0$ and $f_\mathrm{max} = \infty$ have been assumed. When closed-form expressions are unavailable we have a few options. One possibility is to build an ROQ rule for the norm, which requires additional offline computations. Here we consider an alternative. Notice that the norm
\begin{equation} \label{eq:EIMnorm}
\langle h\left(\mb{\lambda}\right) | h\left(\mb{\lambda}\right) \rangle = \sum_{i=1}^m c_i^2
\end{equation}
is expressible in terms of the EIM coefficients $\vec{c} = A^{-1}\vec{h}$. Explicit computation of these coefficients carries an $\bigo{m^2}$ cost, which is larger than the ROQ count of $\bigo{m}$. However, in many applications of interest the waveforms themselves are very expensive to compute and so this cost will still be much smaller than the full likelihood evaluation.

\subsection{ROQ cost and efficiency} \label{sec:ROQcost}

Here we comment on ROQ offline and startup costs as well as the expected speedup for likelilood evaluations. 

To find $m$ basis functions we use the greedy algorithm described in Appendix~\ref{sec:RBapp}. The asymptotic cost of this algorithm applied to a training set with $M$ elements is ${\cal O} \left( N M m \right)$\footnote{To arrive at this scaling note that for hierarchically built spaces $P_i h = P_{i-1} h + \left< e_i|h \right>e_i$. In turn, each inner product computation $\left< e_i|h \right>$ costs ${\cal O} (N)$ when an $N$-point numerical quadrature rule is used.}. Furthermore, the algorithm is trivially parallelized making large $M$ problems accessible. Once the basis is built, an EIM algorithm is used to identify the ROQ points. As described in Appendix~\ref{sec:EIMapp},  the  cost of the EIM is dominated by inversion of a full matrix; in particular the matrix defined in Eq.~(\ref{eq:InterpMatrix}) for the first $i$ basis/points ( see  algorithm 2 in Ref.~\cite{antil2012two} for a equivalent algorithm which utilizes a lower triangular matrix). The asymptotic cost of Alg.~\ref{alg:EIM} and its modified equivalent are ${\cal O} \left( m^4 + N m^2\right)$ and ${\cal O} \left( m^3 + N m^2\right)$ respectively. 

When considering startup costs, we note that the matrix $A$ is data-independent and can be inverted offline. To compute ROQ weights first i) $m$ inner products between the data and all basis are computed from Eq.~\eqref{eq:innerwithbasis} and finally ii) the matrix-vector product \eqref{eq:weights} is performed. Whence the overall startup cost is ${\cal O} \left(mN + m^2 \right)$.

We now compare the cost of full and compressed (ROQ) overlap evaluations, respectively Eq.~\eqref{eq:NumRiem} and Eq.~\eqref{eq:ROQoverlap}. Computational costs stem from evaluating $h(f_i,\mb{\lambda})$ as well as performing the multiplications/sums. When the waveforms are known through closed-form, frequency-domain expressions we expect a speedup factor of approximately $N/(2m)$. For closed-form, time-domain expressions the savings will be even greater if ROQ rule is constructed for Eq.~\eqref{eq:NumRiem} while the EIM interpolant (and hence selected ROQ points) is built in the time-domain. If the waveforms are found by solving ordinary differential equations the speedup is less straightforward to estimate. For example, adaptive time stepping schemes, such as the Runge-Kutta-Fehlberg method, permit large step sizes set by an error threshold (rather than equally spaced samples set by $\Delta f$). Thus, while one should expect fewer ODE steps to evaluate for $m$ (as opposed to $N$) points, the savings would be problem dependent.

%%%%%%%%%%%%%%%%%%%%%%%%%%%%%%%%%%%%%%%%%%%%%
\section{Results}\label{sec:MCMC_ROQ}
%%%%%%%%%%%%%%%%%%%%%%%%%%%%%%%%%%%%%%%%%%%%%

An MCMC algorithm aims to find a chain of $N_{\tt mcmc}$ samples, $\{ {\bf x}_i \}$, that are distributed according to the target probability distribution, $p_{\rm t}({\bf x}_i)$, such that integrals over the probability distribution can be approximated by sums over the points in the chain
\begin{equation}
\int p_{\rm t} ({\bf x}) f({\bf x}) {\rm d}{\bf x} \approx \sum_{i=1}^{N_{\tt mcmc}} f({\bf x}_i).
\end{equation}
The chain of points can be obtained using the Metropolis-Hastings algorithm \cite{annealing2}. The first point, ${\bf x}_1$, is chosen at random from the prior. At iteration $i$ a new point ${\bf y}_{i}$ is drawn from a proposal distribution $q({\bf y}_{i} | {\bf x}_i)$ and the Metropolis-Hastings ratio, $r$, evaluated
\begin{equation}
r = \frac{p_{\rm t}({\bf y}_i) q({\bf x}_{i} | {\bf y}_i)}{p_{\rm t}({\bf x}_i) q({\bf y}_{i} | {\bf x}_i)}.
\end{equation}
A random number $u \in U[0,1]$ is drawn and if $u < r$ the move is accepted, ${\bf x}_{i+1} = {\bf y}_i$; otherwise the move is rejected and ${\bf x}_{i+1}={\bf x}_i$.

In our case, the target distribution is the posterior probability distribution given by Eq.~(\ref{eq:pdf}), which depends on the likelihood and can therefore be approximated using ROQs. To illustrate the method, we will consider the problem of recovering the parameters of a burst signal of the form given in Eq.~(\ref{eq:h_t}) from a noisy data stream.

We include Gaussian white noise with unit power spectral density, $\tilde{S}_n(f) = 1$, and take the parameters of the true signal to be our default ones,  Eq.~(\ref{eq:injected}). We assume that the observation is $32\sec$ long and the data is sampled at $64$Hz. We use a symmetric Gaussian proposal distribution 
$$
q({\bf y}_{i} | {\bf x}_i) \propto \exp\left[-\Gamma_{jk} (x_i^j-y_i^j) (x_i^k-y_i^k)/2\right] \, , 
$$ 
where $\Gamma_{jk} = \langle \partial_j h | \partial_k h \rangle$ is the Fisher information matrix. We use priors on $f_0$ and $\alpha$ that span the range over which the RB and ROQ were built, given by Eq.~(\ref{eq:range_pars}), and priors for the other parameters of $t_c \in [-2,2]$, and $A \in [0.1,10]$. In order to compare the cost and accuracy of the {\it full} (or {\it standard}) MCMC computation vs the ROQ one, we repeat the analysis using the same data, number of MCMC points, proposal distribution and priors, but changing from the full likelihood to the ROQ one. The results are presented in the following sections.

\subsection{Two parameter search}

\begin{table}
\begin{tabular}{l l  c c}
&&\multicolumn{2}{c}{ Recovered Values}\\ 
SNR&\ \ Method&$f_0$&\ \ $\alpha$\\
\hline\hline
\hline\multirow{2}{*}{5}&\ \ Full&\ \ $0.189 \pm 0.095$&\ \ $0.831 \pm 0.194$\\
				&\ \ ROQ&\ \ $0.189 \pm 0.095$&\ \ $0.831 \pm 0.194$\\\hline
\hline\multirow{2}{*}{10}&\ \ Full&\ \ $0.172 \pm 0.081$&\ \ $0.803 \pm 0.136$\\
				 &\ \ ROQ&\ \ $0.172 \pm 0.081$&\ \ $0.803 \pm 0.136$\\\hline
\hline\multirow{2}{*}{20}&\ \ Full&\ \ $0.168 \pm 0.075$&\ \ $0.800 \pm 0.108$\\
				   &\ \ ROQ&\ \ $0.168 \pm 0.075$&\ \ $0.800 \pm 0.108$\\\hline
\hline\multirow{2}{*}{40}&\ \ Full&\ \ $0.212 \pm 0.051$&\ \ $0.872 \pm 0.091$\\
				   &\ \ ROQ&\ \ $0.212 \pm 0.051$&\ \ $0.872 \pm 0.091$\\\hline
\end{tabular}
\caption{\label{twoparamtable}Parameter values recovered, for the waveform frequency $f_0$ and width $\alpha$, using both the full and ROQ likelihoods. Values quoted are the mean and standard deviation estimated from the posterior for a particular noise realisation. The same noise realisation is used for the full and ROQ likelihood calculations for each SNR.}
\end{table}
As a first test we restrict the search to two parameters --- $\{f_0, \alpha\}$ --- while fixing $t_c$ and $A$ to the injected values. In Table~\ref{twoparamtable} we compare the parameter values recovered using the full data set and Riemann sums with those recovered from ROQ likelihoods in one particular noise realization for each of four different SNRs of the injected source. The values are quoted as $\mu_{\rm i} \pm \sigma_{\rm i}$, where the one dimensional marginalised posterior mean, $\mu_i$, and standard deviation, $\sigma_i$, in parameter $i$ are defined from the set of MCMC samples $\{ {\bf x}_j\}$ by
%---------
\begin{equation}
\mu_i = \frac{1}{N_{\tt mcmc}} \sum_{j=1}^{N_{\tt mcmc}} x_j^i, \qquad \sigma_i^2 = \frac{1}{N_{\tt mcmc}-1} \sum_{j=1}^{N_{\tt mcmc}} \left(x_i^j - \mu_i \right)^2 .
\end{equation}
%-----------
In all cases the statistics of the posterior distribution are completely consistent between the full likelihood and ROQ likelihood computations. The only differences are beyond the significant digits quoted in the Table and are much smaller than the corresponding uncertainty in the parameter values arising from noise in the data stream. The ROQ likelihood is extremely accurate, with differences of $10^{-6}$ or smaller, so it is not surprising that the statistical results are indistinguishable.

We can also ask whether the full posterior distributions are consistent between the two likelihoods. This can be achieved by using a Kolmogorov-Smirnov (KS) test~\cite{Press92} to compare the 1D and 2D marginalised posteriors obtained using the two different likelihoods. Figure~\ref{TwoParMargDist} shows the 1D marginalised posteriors for $f_0$ and $\alpha$ computed using the two likelihoods. These are indistinguishable by eye and, more precisely, the $p$-value of the KS test that the distributions agree are $1.0$ (full and ROQ likelihood evaluations agree to within 11 digits) for both $f_0$ and $\alpha$, so there is no evidence of any difference in the recovered posteriors. Again, this is to be expected because of the high accuracy of the ROQ likelihood. 
%%%%
\begin{widetext}
\begin{center} 
\begin{figure}[ht]
\includegraphics[width=0.6\linewidth]{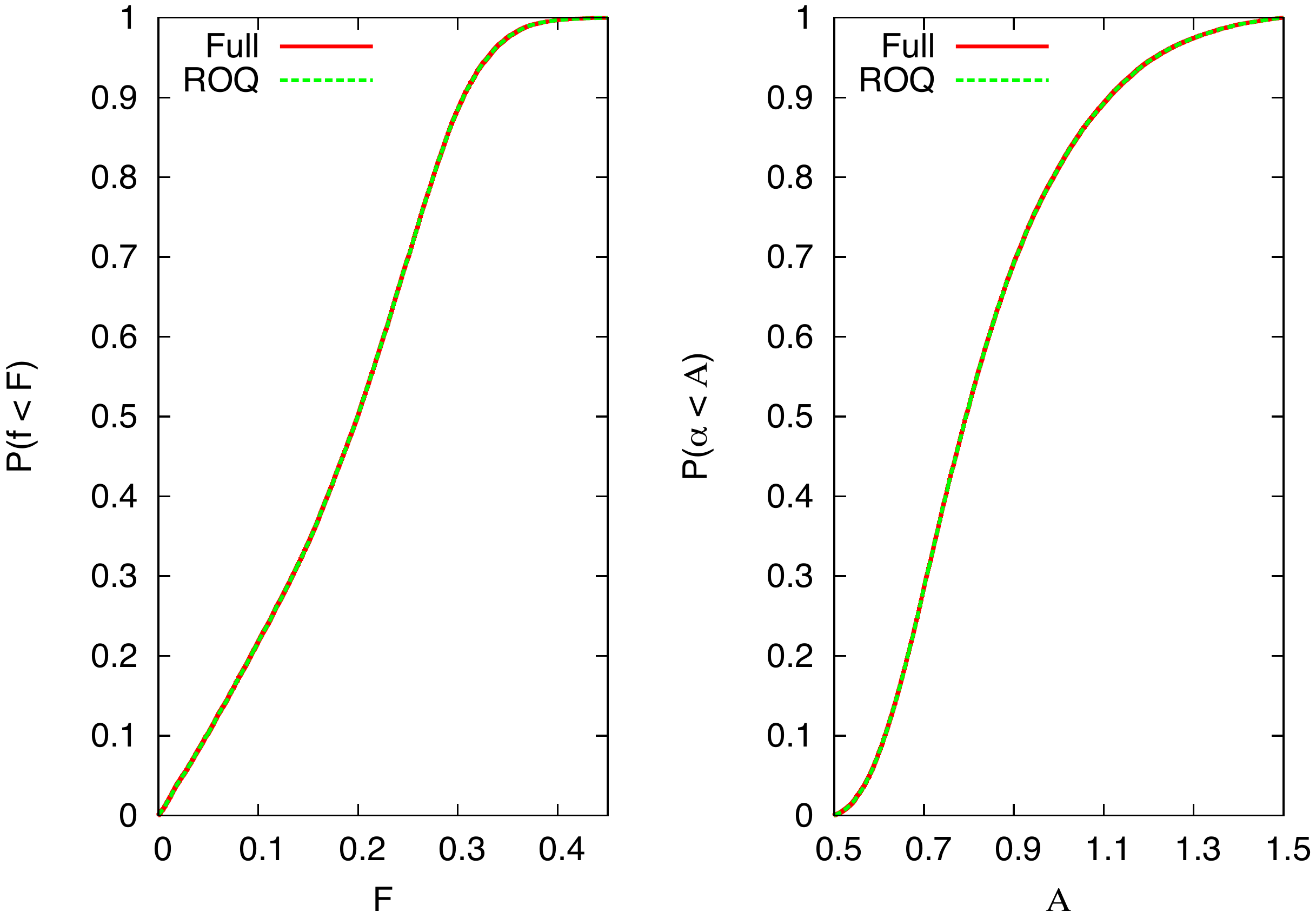}
\caption{\label{TwoParMargDist}Marginalised cumulative probability distributions for $f_0$ (left panel) and $\alpha$ (right panel) for a true source with SNR $\rho=5$. Each panel contains two curves which lie on top of each other, one computed using the Full likelihood and one using the ROQ likelihood. A KS test confirms that the two distributions are the same with probability $1.0$ (full and ROQ likelihood evaluations agree to within 11 digits).}
\end{figure}
\end{center}
\end{widetext}

\subsection{Four parameter search} \label{sec:4d}
We now consider a search over the full four dimensional parameter space $\{ f_0, \alpha, t_c, A\}$. The 1D and 2D marginalised posteriors for a typical noise realisation computed using both the full and ROQ likelihoods are shown in Fig.~\ref{fourparamPDF}, while Table~\ref{fourparamtable} lists the posterior means and standard deviations found in a particular noise realisation using both techniques for a variety of SNRs of the true source. As in the two parameter case, we find that the statistics derived from the posterior distributions (e.g., the mean, standard deviation, quantiles, etc.) are completely consistent between the full and ROQ likelihoods and, more precisely, $p$-values of the marginalized distributions are $\sim0.25$ -- $0.75$ for $10^3$ point MCMC chains. The KS statistic, which measures the maximum difference in the full and ROQ cumulative probability distributions, computed from the marginalized posteriors were $\sim 10^{-2}$. As described in Sec.~\ref{sec:ext_par} these small differences stem from applying an ROQ rule built for $t_c=0$ to non-zero values of $t_c$ (see Fig.~\ref{fig:errorstc}). While the resulting errors are smaller than the typical width of the posterior, if higher accuracy is desired the alternative approaches discussed in Section.~\ref{sec:ext_par} can be used.

\begin{widetext}
\begin{center}
\begin{figure}[ht]
\begin{tabular}{cc}
\includegraphics[width=0.5\textwidth]{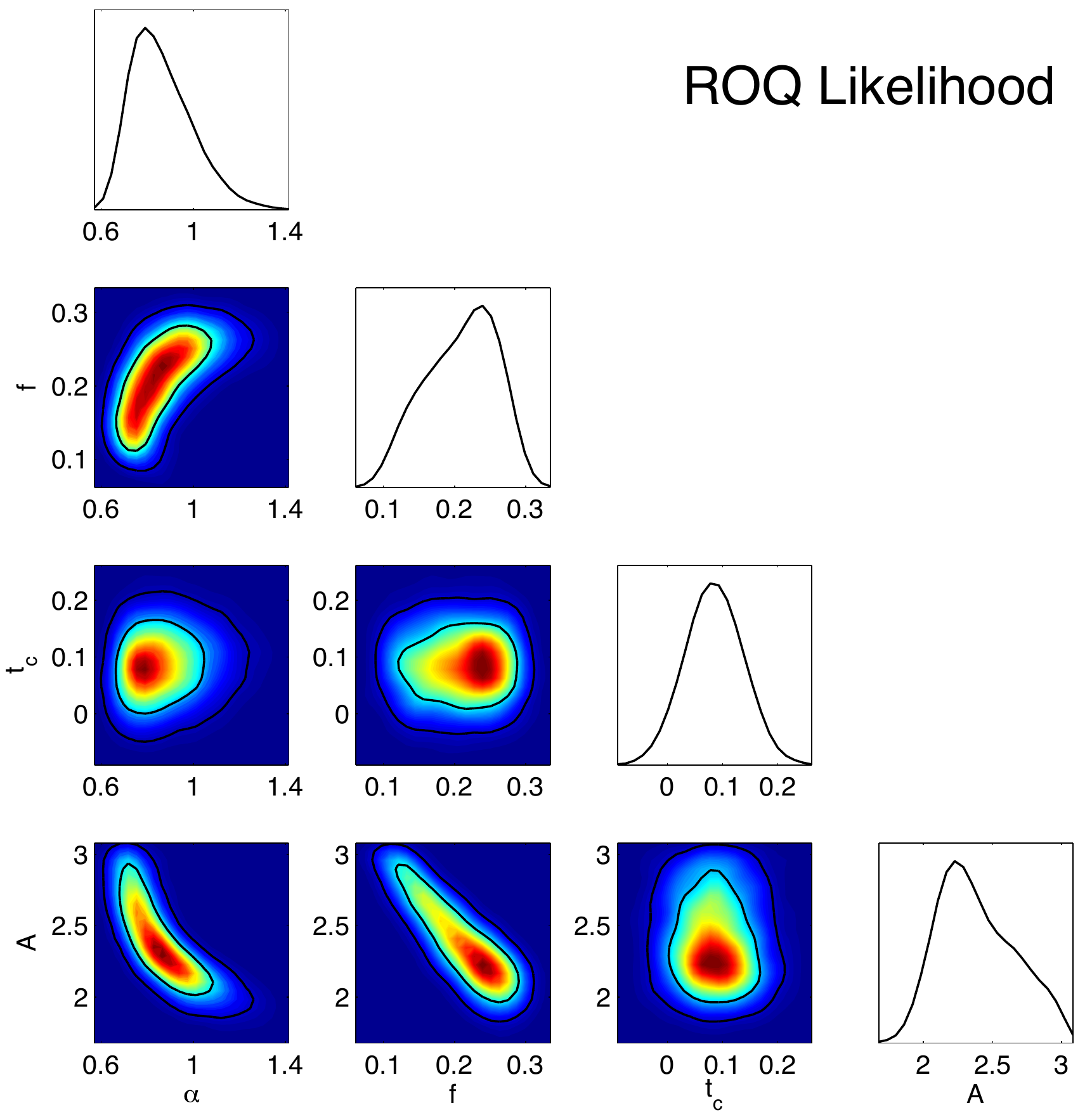}&
\includegraphics[width=0.5\textwidth]{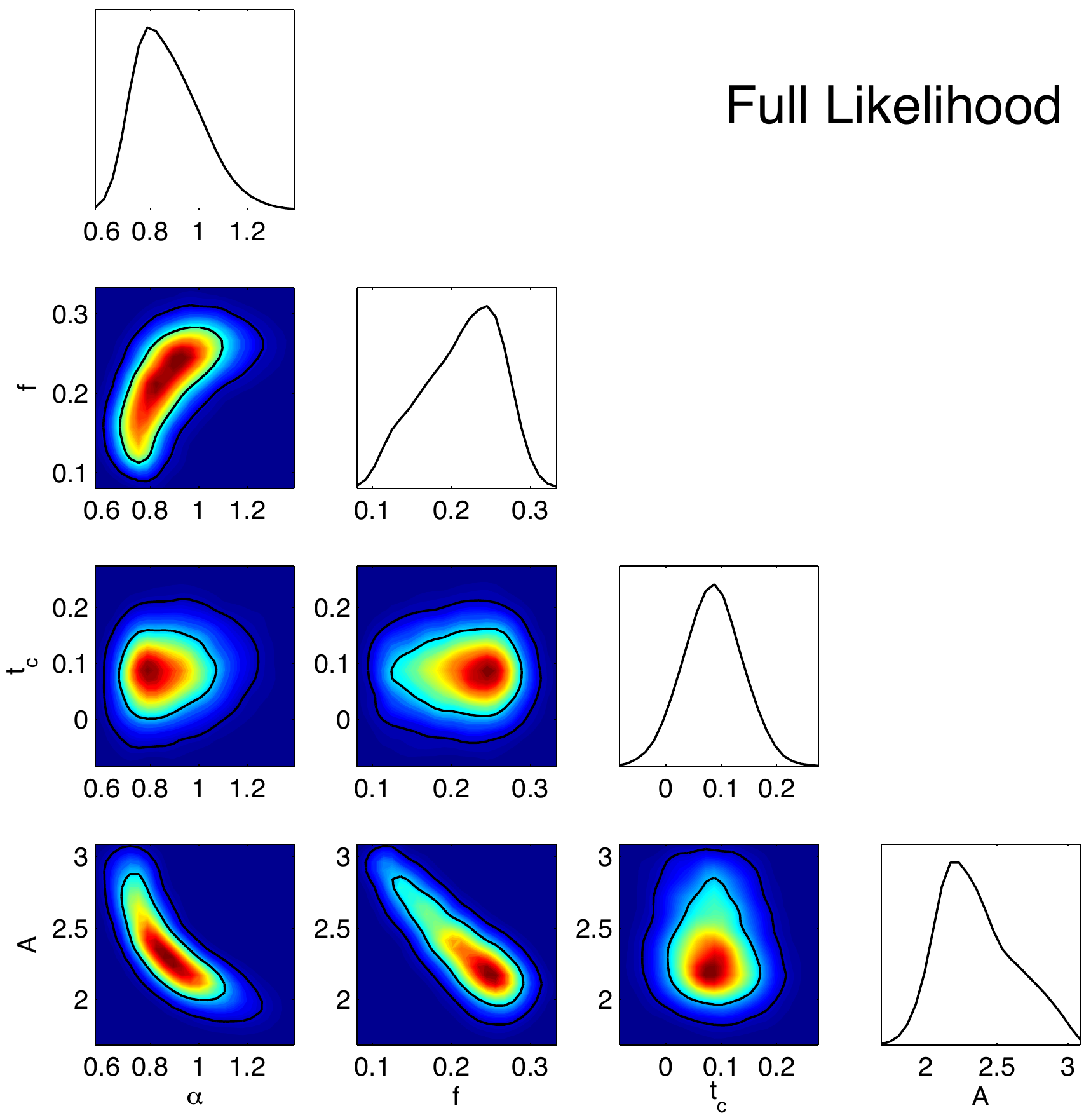}
\end{tabular}
\caption{\label{fourparamPDF}PDFs obtained for an injected source with SNR $\rho= 10$, employing standard and ROQ MCMC computations in a four-parameter space, namely $A$, $t_c$, $f_0$ and $\alpha$. The figures qualitatively show the agreement between the two techniques, see Sec.~\ref{sec:4d} for more details.}
\end{figure}
\end{center}
\end{widetext}

\begin{table*}
\begin{tabular}{l l c c c c}
&&\multicolumn{4}{c}{Recovered values}\\
SNR&Method&\ \ $f_0$&$\alpha$&$t_c$&$A$\\
\hline\multirow{2}{*}{5}&Full& \ \ $0.217 \pm 0.069$&\ \ $0.896 \pm 0.194$&\ \ $0.068 \pm 0.104$&\ \ $1.704 \pm 0.379$\\
				&ROQ&\ \ $0.217 \pm 0.068$&\ \  $0.897 \pm 0.196$&\ \ $0.069 \pm 0.104$&\ \ $1.702 \pm 0.375$\\\hline
\hline\multirow{2}{*}{10}&Full&\ \  $0.212 \pm 0.048$&\ \ $0.875 \pm 0.132$&\ \ $0.084 \pm 0.053$&\ \ $2.362 \pm 0.278$\\
				 &ROQ&\ \ $0.209 \pm 0.050$&\ \ $0.866 \pm 0.132$&\ \ $0.085 \pm 0.052$&\ \ $2.387 \pm 0.287$\\\hline
\hline\multirow{2}{*}{20}&Full&\ \ $0.225 \pm 0.029$&\ \ $0.891 \pm 0.093$&\ \ $0.092 \pm 0.028$&\ \ $2.944 \pm 0.176$\\
				&ROQ&\ \ $0.224 \pm 0.029$&\ \ $0.892 \pm 0.093$&\ \ $0.093 \pm 0.028$&\ \ $2.944 \pm 0.177$\\\hline
\hline\multirow{2}{*}{40}&Full&\ \ $0.248 \pm 0.009$&\ \ $0.981 \pm 0.041$&\ \ $0.097 \pm 0.016$&\ \ $3.471 \pm 0.157$\\
			           &ROQ&\ \ $0.248 \pm 0.009$&\ \ $0.981 \pm 0.042$&\ \ $0.097 \pm 0.016$&\ \ $3.471 \pm 0.157$\\\hline
\end{tabular}
\caption{\label{fourparamtable}As Table~\ref{twoparamtable} but for searches over the full set of four parameters: waveform frequency $f_0$ and width $\alpha$, coalescence time $t_c$ and amplitude $A$. The parameter valuesare recovered using the full and ROQ likelihoods. Values quoted are the mean and standard deviation estimated from the posterior for a particular noise realisation. The same noise realisation is used for the full and ROQ likelihood calculations for each SNR}
\end{table*}

Having established the equivalence of the results for the full and ROQ likelihoods, we can now compare the run time. The ROQ likelihood has a higher initial cost, since the data-specific weights (\ref{eq:weights}) have to be computed prior to beginning the MCMC. In general this start-up cost is a tiny fraction of the total run time of the MCMC algorithm\footnote{This already small cost can be further reduced by inverting the matrix $A$ Eq.~\eqref{eq:InterpMatrix} offline, see Sec.~\ref{sec:ROQcost}.}. For the burst waveforms used in this paper, the total time taken to compute the weights is $\sim 10$ms, which is comparable with $\sim 85$ MCMC chains using the full likelihood. By comparison a resolved MCMC simulation, for example the one leading to table~\ref{fourparamtable}, requires $\sim 5 \times 10^5$ MCMC chains. Evidently, for this problem, the start-up time is a negligible fraction $0.01\%$ of overall cost for a resolved MCMC simulation using the full likelihood. In light of the scalings described in Sec.~\ref{sec:ROQcost} we expect negligible start-up costs whenever $m < N/2$.

\begin{figure}[ht] \label{fig:mcmc_time}
\includegraphics[width=0.98 \linewidth]{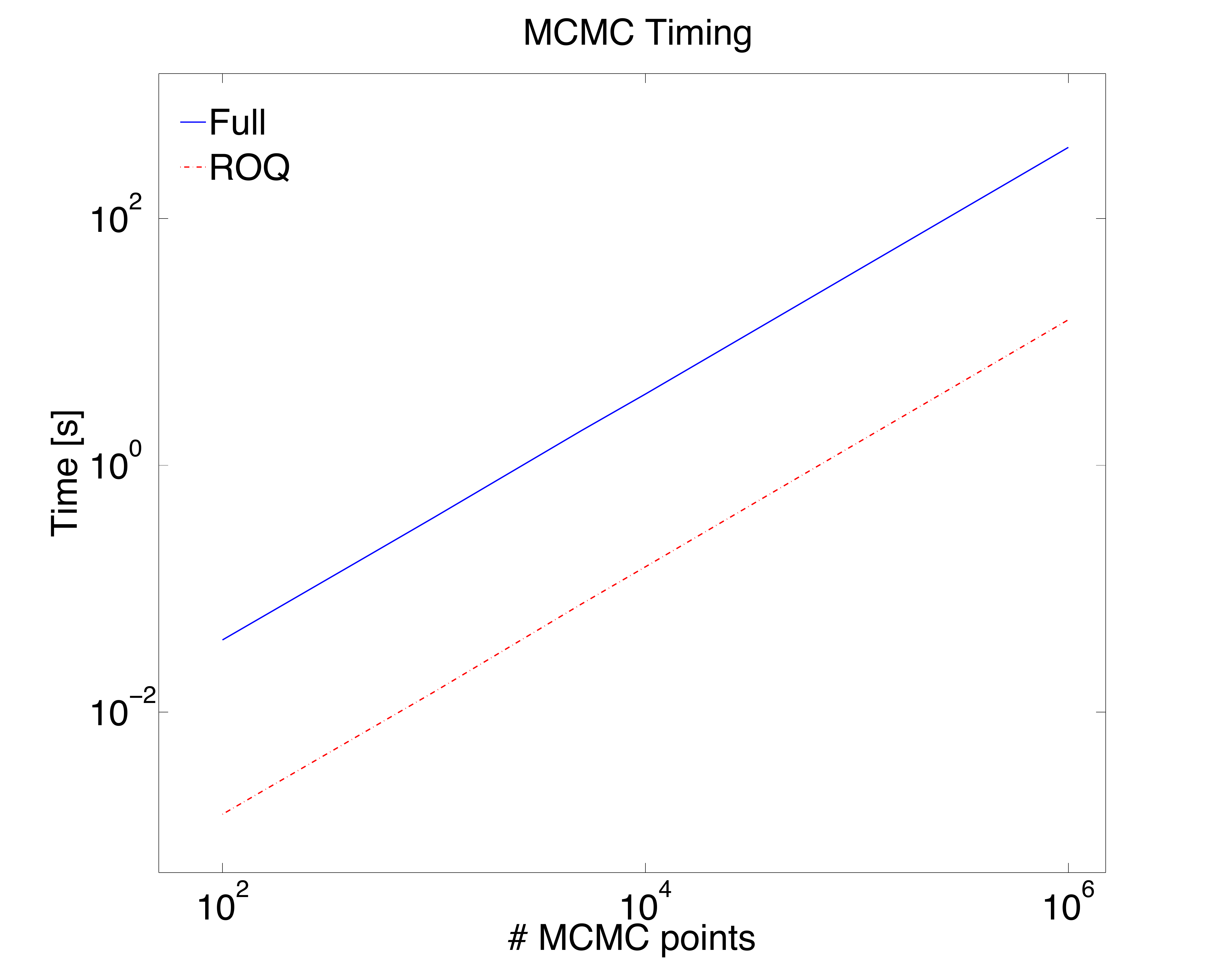}
\caption{\label{MCMCruntime}Runtime as a function of MCMC chains for $N_{\text{ \tiny{mcmc}}}=10^6$ samples. The red (dotted) line shows the timing for ROQ computations and the blue (dashed) line the timing for the standard MCMC computations.}
\end{figure}

\begin{figure}[ht]
\includegraphics[width=1.1 \linewidth]{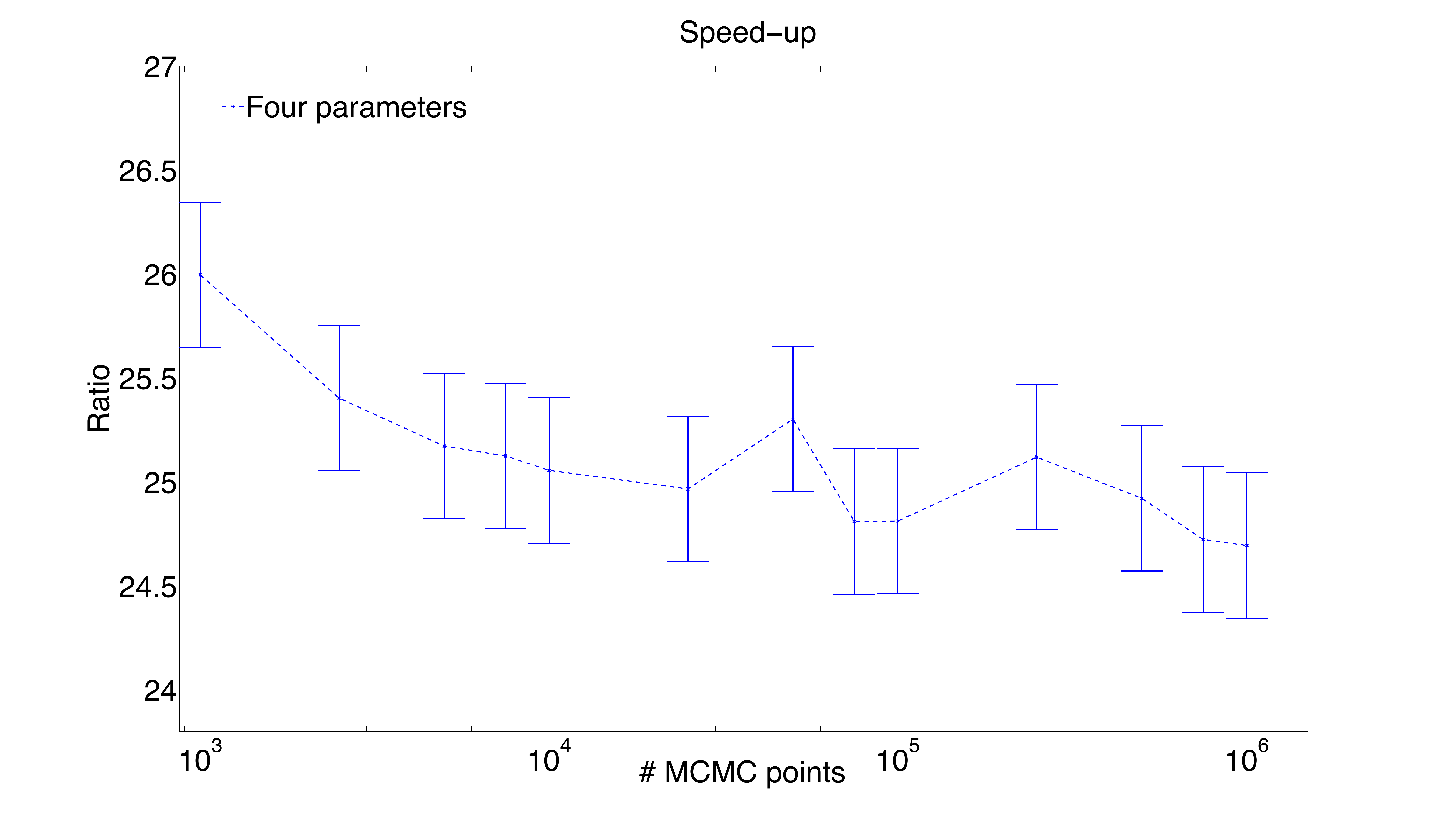}
\caption{\label{MCMCtimeratio}Time ratio (speed-up) of MCMC simulations using a standard quadrature rule and the ROQ one. The figure shows the mean of the speed-up obtained by performing simulations with different seed values for the MCMC. See the Sec.~\ref{sec:4d} for details.}
\end{figure}

In Fig.~\ref{MCMCruntime} we show the time taken to run the MCMC search, i.e., after the initial set-up time, using the full and ROQ likelihoods. As we can see the ROQ is two orders of magnitude faster that the full likelihood computation. Figure~\ref{MCMCtimeratio} shows the ratio of the runtimes for the ROQ and full searches. The speed-up is seen to be $\sim 25$, which is expected in light of the scalings given in Sec.~\ref{sec:ROQcost}.

The cost of the MCMC search grows linearly with the number of MCMC points, as we would expect, since the run-time is determined primarily by the cost of likelihood evaluations. The speed-up from using the ROQ is, in this case, a factor of $\sim25$. This factor will of course be problem and implementation-dependent,  but it is roughly the ratio between the total number of frequency samples $N/2$ and the number of ROQ subsamples $m$. This ratio will depend on various aspects of the problem --- the sampling cadence, total observation time, the allowed range for the parameters, and the waveform model itself. For example, if we know in advance the frequency and duration of the burst then carefully choosing a sampling rate and observation time just large enough for the source used in this paper reduces the speed-up to $\sim 10$. Such tuning of the cadence and observation time is effectively a compression of the likelihood, and is very effective for a simple model of this type. The fact that even after such tuning the ROQ rule can show a significant speed-up illustrates the power of the method. In other problems, speed-up factors of $10$--$100$ are typical and factors of $1000$ are possible, but these have to be computed on a case by case basis and will be reported elsewhere. An investigation of the speed-ups for inspiral waveforms is currently underway. 

%%%%%%%%%%%%%%%%%%%%%%%%%%%%%
\section{Summary}  \label{sec:conc}
%%%%%%%%%%%%%%%%%%%%%%%%%%%%%

In this paper we have proposed using a modification of the Reduced Order Quadratures (ROQ) of Ref.~\cite{antil2012two} for fast, accurate evaluations of the correlation between a given data stream and a family of gravitational waveforms. The modification is designed for Markov chain Monte Carlo (MCMC) parameter estimation studies and as such, it is adapted to a particular stream of (noisy) data. The resulting speed-up is not at the expense of reduced accuracy but, instead, Reduced Order Modeling is used to build application and data-specific quadratures for the problem at hand. 

The ROQ rule requires an offline computation to build a waveform basis and identify a distribution of sparse data samples. This application-specific information can be stored to file and reused for any stream of data. Then, for a given data set we compute data-specific weights using Eq.~\eqref{eq:weights}; the overall cost of this computation is negligible. Fast and accurate compressed likelihood computations are then performed with Eq.~\eqref{eq:ROQoverlap}, which can be implemented within existing MCMC codes in a non-intrusive manner.

For the particular application considered here as an illustration of the concept, models of burst gravitational waves, we have found speedups of $\sim \times 25$, depending on settings such as the central frequency of the wave, the damping factor, observation period, and sampling rate. These speedups are expected to increase with the complexity and fidelity of the model. 

In Ref.~\cite{PhysRevD.86.084046} it was found that the number of Reduced Basis waveforms needed to represent the space of inspiral waveforms in the post-Newtonian stationary phase approximation barely increases when (non-precessing) spins are taken into account. Since ROQ by design uses the same number of nodal points as the number of basis functions needed to represent the space of waveforms within a given accuracy, the approach holds the promise of beating the curse of dimensionality. 
There is also evidence that the case of precessing binaries is amenable to dimensional reduction \cite{Galley:2010rc}.

%%%%%%%%%%%%%%%%%%%%%%%%%%%%%
\section{Acknowledgments} 
%%%%%%%%%%%%%%%%%%%%%%%%%%%%%
This work was supported in part by NSF Grants PHY1208861 and PHY1005632 to the University of Maryland. PC's work is supported by a Marie Curie Intra-European Fellowship within the 7th European Community Framework Programme (PIEF-GA-2011-299190), and thanks the University of Maryland for hospitality while part of this work was completed. SEF acknowledges support from the Joint Space Science Institute. JG's work is supported by the Royal Society. MT thanks the National Institute for Theoretical Physics (NiTheP) at Stellenbosch University, South Africa, where part of this work was done, for its hospitality. We thank Chad Galley and Jason Kaye for helpful discussions and comments on the manuscript. We also thank Harbir Antil, Collin Capano, Philip Graff, Frank Herrmann, Tyson Littenberg, Ilya Mandel, Evan Ochsner,   Ricardo Nochetto and Rory Smith for helpful and insightful discussions.

\appendix
%%%%%%%%%%%%%%%%%%%%%%%%%%%%%
\section{Reduced Basis} \label{sec:RBapp}
%%%%%%%%%%%%%%%%%%%%%%%%%%%%%
In its simplest form, such as when the waveforms are inexpensive to compute, the greedy algorithm for building a RBs has as input a set of parameter values 
\be
{\cal T}:= \{\mlam_i \}_{i=1}^M \label{eq:training}
\ee
usually called {\it training points}, and associated waveforms $ \{h(\cdot; \mlam_i ) \}_{i=1}^M$, usually called the {\it training set}. 

Part of the output is a hierarchical set of parameter values $\{ \mlam_1, \mlam_2, \cdots , \mlam_m \} \subseteq {\cal T}$ (with $m\leq M$, and $m< M$ or even $m \ll M$ if the problem is amenable to dimensional reduction) called the greedy points, and associated waveforms, which constitute the RBs,  
\be
\text{RB}:= \{ e_1 (\cdot):= h(\cdot, \mlam_1), \cdots, e_m (\cdot):= h(\cdot, \mlam_m)\} \, . \label{eq:RB}
\ee
The RB serves as a representation of the waveforms in the training set and, if the latter is dense enough, of the whole continuum. The optimal representation by a basis is known to be the orthogonal projection ${\cal P}_m$ onto its span. This result is a standard linear algebra one, independent of Reduced Basis or Reduced Order Modeling. That is, the approximation   
\be
h (\cdot; \mlam) \approx \sum_{i=1}^m c_i (\mlam ) e_i (\cdot )   \label{eq:RBapprox}
\ee
minimizes the error,  
$$
\left \| h (\cdot; \mlam) - \sum_{i=1}^m c_i (\mlam) e_i (\cdot) \right \|^2\,  ,
$$
when the coefficients $c_i$ are chosen such that
\be
\left \langle h (\cdot; \mlam) -  \sum_{i=1}^m c_i (\mlam) e_i (\cdot) \middle| e_j (\cdot ) \right \rangle = 0 \quad \forall \quad e_j \in \text{RB} \, . \label{eq:LS} 
\ee
The solution to (\ref{eq:LS}) is 
\be
c_i (\mlam )= \sum_{j=1}^m(G^{-1} )_{ij} \langle h (\cdot ; \mlam ) | e_j (\cdot ) \rangle \, , \label{eq:optLS}
\ee
where $G^{-1}$ is the inverse of the {\it Grammian} or {\em Gram matrix} $G$, with entries
$$
G_{ij} := \langle e_i  | e_j \rangle \, .  
$$
If the basis is orthonormal, this matrix is the identity and one recovers the familiar expression 
$$
h \approx \sum_{i=1}^m \langle h | e_i \rangle e_i \, . 
$$
In general the RB waveforms selected by the greedy algorithm will not be orthonormal. Then at each greedy iteration one can use a Gram-Schmidt (GS) procedure to orthonormalize the RB or, equivalently, simply invert the Gram matrix. In either case, for any given basis, the optimal approximation of the form (\ref{eq:RBapprox}) is given by 
\be
h (\cdot; \mlam) \approx {\cal P}_m h (\cdot ; \mlam) := \sum_{i=1}^m c_i (\mlam ) e_i (\cdot ) \, ,  \label{eq:Pm}
\ee
with the coefficients $c_i$ given by Eq.~(\ref{eq:optLS}). Notice that since the approximant (\ref{eq:Pm}) is defined in a completely geometric way, as the orthogonal projection onto the {\em span} of the RB elements, it is independent of whether a GS procedure is carried out or not. The RB (\ref{eq:RB}), at the same time, is composed of a set of the ``most relevant'' {\em physical waveforms}. 

The precise algorithm to choose the greedy points is described in Alg.~\ref{alg:Greedy}. Given an arbitrary user-defined tolerance error $\epsilon$, the algorithm stops when the approximation (\ref{eq:Pm}) meets the tolerance,
$$
\| h(\cdot ; \mlam ) - {\cal P}_m h(\cdot ; \mlam ) \|^2 \leq \epsilon \,\,\,\, \forall \,\, \mlam \in {\cal T}.  
$$
In all expressions the scalar product $\langle \cdot |  \cdot \rangle$ and its associated norm might be weighted. In the context of GW physics a natural choice is that one given by Eq.~(\ref{eq:scalarprod}), but any other choice is possible. 

\hspace{0.5cm}

{\scriptsize
\begin{algorithm}[H]
\caption{Brief description of the Greedy Algorithm}
\label{alg:Greedy}
\begin{algorithmic}[1]
\State {\bf Input:} $ \{ \mlam_i \, , h(\cdot; \mlam_i) \}_{i=1}^M$,  $\epsilon$ 
\vskip 10pt
\State {\bf Seed choice} (arbitrary):  $\mlam_1$
\State RB = $\{ h (\cdot ; \mlam_1) \}$
\State $i=1$ and $\sigma_1 = 1$
\While{$\sigma_i \ge \epsilon$}
\State $i=i+1$
\State $\sigma_i = \max_{ \mlam \in {\cal T} }\| h(\cdot; \mlam ) - {\cal P}_{(i-1)} h(\cdot; \mlam ) \|^2$ \label{GreedyErrs}
\State $\mlam_{i} = \text{argmax}_{ \mlam \in {\cal T} }\| h(\cdot; \mlam ) - {\cal P}_{(i-1)} h(\cdot; \mlam ) \|^2$ 
\State RB = RB $\cup \, h(\cdot, \mlam_i)$
\EndWhile
\vskip 10pt
\State {\bf Output:} RB and greedy points
\end{algorithmic}
\end{algorithm}
}

%%%%%%%%%%%%%%%%%%%%%%%%%%%%%
\section{The Empirical Interpolation Method} \label{sec:EIMapp}
%%%%%%%%%%%%%%%%%%%%%%%%%%%%%

The EIM approach is very different, in goals and scope, to any variation of standard polynomial interpolation, which was described for completeness in Sec.~\ref{sec:interpolation}. The goal of EIM is to deal with parametrized problems characterized by non-polynomial bases. The set of EIM points is nested and hierarchical, as one would want when solving differential equations, and easily handles unstructured meshes in several dimensions. 

Consider a basis $\{ e_i (x) \}_{i=1}^m$ whose span accurately approximates the functions $h(x; \mlam)$. For definiteness we will denote by $x$ the physical dimension(s) and $\mlam$ the parametrization of these functions.
For example, if $h$ is a GW then $x$ could denote time or frequency, and $\mlam $ the intrinsic or extrinsic parameters of the system. Let $\{x_i\}_{i=1}^N$ denote a set of $N$ points and define the corresponding $N$-vector $\vec{x} = \left(x_1, x_2, \dots, x_N \right)^{T}$. Discrete objects arise from evaluating continuous functions at $\vec{x}$. For example, defining $h_i(\mlam) = h(x_i; \mlam)$, the GW $N$-vector is $\vec{h}(\mlam) = h(\vec{x}; \mlam)$. Similarly, $\vec{e}_i = e_i(\vec{x})$ denotes the $i^{th}$ basis function evaluated at $\vec{x}$.

Given an input of $m$ {\em evaluated} basis functions $\{ \vec{e}_i \}_{i=1}^m$ the output of the EIM algorithm is a set of $m$ EIM points 
\begin{align}
\{ X_i \}_{i=1}^m \subset \{x_i\}_{i=1}^N \label{eq:EIM_points}
\end{align}
selected as a subset of $\{x_i\}_{i=1}^N$. If a function $ h (x ; \mlam)$ is known at the EIM points $\{ X_i \}_{i=1}^m$, the EIM interpolant can predict with high accuracy the function at any other value of $\{x_i\}_{i=1}^N$. It is an interpolant in the usual sense, meaning that it agrees with the interpolated function at the interpolation points,   
$$
{\cal I }_m [h] (X_i, \mlam) = h(X_i, \mlam) \quad \text{for } i=1,\ldots, m \, .
$$ 
The EIM interpolant is given by Eq.~(\ref{eq:DEIM}), while the selection of the EIM points is described in Algorithm \ref{alg:EIM}. To assist with the description of the EIM algorithm we define the $j$-term empirical interpolant built from the first $j$ basis functions and points
\be
{\cal I}_j [h](x;\mb{\lambda}) := \sum_{i=1}^j c_i (\mlam) e_i(x)\, ,  \label{eq:interpdefI}
\ee
where the $c_i$ coefficients are solutions to the $j$-point interpolation problem 
\begin{align}
{\cal I}_j [h](X_k;\mb{\lambda}) = h(X_k;\mb{\lambda}) , \qquad \forall \, k=1,\dots,j .
\end{align}

\hspace{0.5cm}

{\scriptsize
\begin{algorithm}[H]
\caption{Selection of EIM Points}
\label{alg:EIM}
\begin{algorithmic}[1]
\State {\bf Input:} Evaluated basis $\{ \vec{e}_i \}_{i=1}^m$ and points $\{x\}_{i=1}^N$
\vskip 10pt
\State $i = \text{argmax} | \vec{e}_1 |$ {\bf Comment:} here \text{argmax} takes a vector and returns the index of its largest entry. 
\State Set $X_1 = x_i$
\vskip 10pt
\For{$j = 2 \to m$} 
\State Find ${\cal I}_{j-1} [e_j](\vec{x})$
\State Compute the point-wise error $\vec{r} = {\cal I}_{j-1} [e_j](\vec{x}) - \vec{e}_j$
\State $i = \text{argmax} |\vec{r}|$
\State Set $X_j = x_i$
\EndFor
\vskip 10pt
\State {\bf Output:} EIM points $\{ X_i \}_{i=1}^m$
\end{algorithmic}
\end{algorithm}
}

\noindent {\em Comments} \\
{\bf 1.} In standard polynomial interpolation the interpolant is a linear combination of polynomials and function values, as in Eq.~(\ref{eq:interp}). In the EIM the interpolant is a linear combination of (in the case of interest for this paper), waveforms and function values in the physical dimension(s), as given more precisely by Eq.~(\ref{eq:interpdef}). Parametrization and ``physical'' dimensions play a dual role. 
\\
\noindent {\bf 2.} Unlike Gaussian (e.g., Chebyshev) interpolation nodes, EIM nodes are nested and hierarchical. Given a hierarchical basis
$$
%\left \{ e_1({\cdot}) \right \} \subset \left \{ e_1({\cdot}), e_2({\cdot}) \right \} \subset \ldots \subset  \{ e_i(\cdot ) \}_{i=1}^m
\left \{ e_1(x) \right \} \subset \left \{ e_1(x), e_2(x) \right \} \subset \ldots \subset  \{ e_i(x) \}_{i=1}^m
$$
an associated set of EIM points
$$
\{ X_1 \} \subset  \{ X_1, X_2 \} \subset  \ldots \subset  \{ X_i \}_{i=1}^m 
$$
is defined. Each set of $p$ EIM nodes is included within the set of $p'$ EIM nodes whenever $p < p'$ and only depends on the basis of dimension $p$. 
\\
\noindent {\bf 3.} The empirical interpolant satisfies
\begin{align*}
\max_{\mlam} \left\| h(\cdot ;\mb{\lambda}) - {\cal I}_m [h(\cdot ;\mb{\lambda})] \right\|^2 \leq \Lambda_m^2 \sigma_m \, ,
\end{align*}
where $\sigma_m$ characterizes the representation error of the basis as defined in Eq.~\eqref{eq:greedyErr} and $\Lambda_m$ is a computable Lebesgue constant (see Theorem 2 of Ref.~\cite{antil2012two}). Furthermore, due to the slow growth of $\Lambda_m$, often comparable to the best possible scaling~\cite{Maday_2009}, the interpolant is said to be nearly optimal.

\bibliographystyle{physrev}
\bibliography{references}

\begin{thebibliography}{10}

\bibitem{LIGO_web}
LIGO - http://www.ligo.caltech.edu/.

\bibitem{VIRGO_web}
Virgo - https://wwwcascina.virgo.infn.it.

\bibitem{GEO_web}
GEO600 - http://www.geo600.uni-hannover.de/.

\bibitem{KAGRA_web}
KAGRA - http://gwcenter.icrr.u-tokyo.ac.jp/en/.

\bibitem{Abadie:2010cfa}
LIGO Scientific, J.~Abadie {\em et~al.},
\newblock Class. Quantum Grav. {\bf 27}, 173001 (2010), arXiv:1003.2480.
%%CITATION = 1003.2480;%%

\bibitem{Brown:2012gs}
D.~A. Brown, A.~Lundgren, and R.~O'Shaughnessy,
\newblock (2012), arXiv:1203.6060.

\bibitem{Ajith:2011ec}
P.~Ajith,
\newblock Phys. Rev. {\bf D84}, 084037 (2011), arXiv:1107.1267.

\bibitem{Centrella:2010zf}
J.~M. Centrella, J.~G. Baker, B.~J. Kelly, and J.~R. van Meter,
\newblock Ann.Rev.Nucl.Part.Sci. {\bf 60}, 75 (2010), arXiv:1010.2165.
%%CITATION = ARXIV:1010.2165;%%

\bibitem{Buonanno_etal:PRD70}
A.~Buonanno, Y.~Chen, Y.~Pan, and M.~Vallisneri,
\newblock Phys. Rev. D {\bf 70}, 104003 (2004).

\bibitem{Buonanno:2002fy}
A.~Buonanno, Y.-b. Chen, and M.~Vallisneri,
\newblock Phys. Rev. {\bf D67}, 104025 (2003), arXiv:gr-qc/0211087.
%%CITATION = GR-QC/0211087;%%

\bibitem{Pan_etal:PRD69}
Y.~Pan, A.~Buonanno, Y.~Chen, and M.~Vallisneri,
\newblock Phys. Rev. D {\bf 69}, 104017 (2004).

\bibitem{Littenberg:2012uj}
T.~B. Littenberg, J.~G. Baker, A.~Buonanno, and B.~J. Kelly,
\newblock (2012), arXiv:1210.0893.
%%CITATION = ARXIV:1210.0893;%%

\bibitem{Chatziioannou:2012rf}
K.~Chatziioannou, N.~Yunes, and N.~Cornish,
\newblock (2012), arXiv:1204.2585.
%%CITATION = ARXIV:1204.2585;%%

\bibitem{Yunes:2009ke}
N.~Yunes and F.~Pretorius,
\newblock Phys.Rev. {\bf D80}, 122003 (2009), arXiv:0909.3328.
%%CITATION = ARXIV:0909.3328;%%

\bibitem{Canizares:2012is}
P.~Canizares, J.~R. Gair, and C.~F. Sopuerta,
\newblock Phys.Rev. {\bf D86}, 044010 (2012), arXiv:1205.1253.
%%CITATION = ARXIV:1205.1253;%%

\bibitem{Vallisneri:2011ts}
M.~Vallisneri,
\newblock Phys. Rev. Lett. {\bf 107}, 191104 (2011), arXiv:1108.1158.
%%CITATION = ARXIV:1108.1158;%%

\bibitem{Cornish:2010kf}
N.~J. Cornish,
\newblock (2010), arXiv:1007.4820.
%%CITATION = ARXIV:1007.4820;%%

\bibitem{Mitra:2005gm}
S.~Mitra, S.~V. Dhurandhar, and L.~S. Finn,
\newblock Phys. Rev. {\bf D72}, 102001 (2005), arXiv:gr-qc/0507011.
%%CITATION = GR-QC/0507011;%%

\bibitem{brown_sc_2013_13}
J. Kaye, {\em The interpolation of gravitational waveforms}, Thesis, Brown
  University, 2012, www.dam.brown.edu/scicomp/reports/2013-8/.

\bibitem{Cannon:2012gq}
K.~Cannon, J.~Emberson, C.~Hanna, D.~Keppel, and H.~Pfeiffer,
\newblock (2012), arXiv:1211.7095.
%%CITATION = ARXIV:1211.7095;%%

\bibitem{Smith:2012du}
R.~Smith, K.~Cannon, C.~Hanna, D.~Keppel, and I.~Mandel,
\newblock (2012), arXiv:1211.1254.
%%CITATION = ARXIV:1211.1254;%%

\bibitem{graff2012bambi}
P.~Graff, F.~Feroz, M.~Hobson, and A.~Lasenby,
\newblock Monthly Notices of the Royal Astronomical Society {\bf 421}, 169
  (2012).

\bibitem{antil2012two}
H.~Antil, S.~E. Field, F.~Herrmann, R.~H. Nochetto, and M.~Tiglio,
\newblock Journal of Scientific Computing , 1 (2013), arXiv:1210.0577 [cs.NA].

\bibitem{Owen:1995tm}
B.~J. Owen,
\newblock Phys. Rev. {\bf D53}, 6749 (1996), arXiv:gr-qc/9511032.

\bibitem{Babak:2009ua}
S.~Babak, J.~R. Gair, and E.~K. Porter,
\newblock Class.Quant.Grav. {\bf 26}, 135004 (2009), arXiv:0902.4133.
%%CITATION = ARXIV:0902.4133;%%

\bibitem{Allen:2005fk}
B.~Allen, W.~G. Anderson, P.~R. Brady, D.~A. Brown, and J.~D. Creighton,
\newblock (2005), arXiv:gr-qc/0509116.
%%CITATION = GR-QC/0509116;%%

\bibitem{Barrault2004667}
M.~Barrault, Y.~Maday, N.~C. Nguyen, and A.~T. Patera,
\newblock Comptes Rendus Mathematique {\bf 339}, 667 (2004).

\bibitem{Maday_2009}
Y.~Maday, N.~C. Nguyen, A.~T. Patera, and S.~H. Pau,
\newblock Communications on Pure and Applied Analysis {\bf 8}, 383 (2009).

\bibitem{Abadie:2012rq}
LIGO Scientific Collaboration, Virgo Collaboration, J.~Abadie {\em et~al.},
\newblock Phys.Rev. {\bf D85}, 122007 (2012), arXiv:1202.2788.
%%CITATION = ARXIV:1202.2788;%%

\bibitem{Pinnau2008}
R.~Pinnau,
\newblock Model reduction via proper orthogonal decomposition,
\newblock in {\em Model Order Reduction: Theory, Research Aspects and
  Applications}, edited by W.~H. A.Schilders, H.~A. van~der Vorst, and
  J.~Rommes, , Mathematics in Industry Vol.~13, pp. 95--109, Springer Berlin
  Heidelberg, 2008.

\bibitem{Stewart:1993:EHS:166597.166599}
G.~W. Stewart,
\newblock SIAM Rev. {\bf 35}, 551 (1993).

\bibitem{Maday:2002:PCT:608985.609022}
Y.~Maday, A.~T. Patera, and G.~Turinici,
\newblock J. Sci. Comput. {\bf 17}, 437 (2002).

\bibitem{Veroy2003619}
K.~Veroy, C.~Prud'homme, and A.~T. Patera,
\newblock Comptes Rendus Mathematique {\bf 337}, 619  (2003).

\bibitem{prud'homme:70}
C.~Prud'homme {\em et~al.},
\newblock Journal of Fluids Engineering {\bf 124}, 70 (2002).

\bibitem{FLD:FLD867}
K.~Veroy and A.~T. Patera,
\newblock International Journal of Numerical Methods in Fluids {\bf 47}, 773
  (2005).

\bibitem{PateraReview}
A.~Patera and G.~Rozza,
\newblock Arch. Comput. Methods Eng. {\bf 15}, 229 (2008).

\bibitem{Nguyen_2009}
N.-C. Nguyen, G.~Rozza, and A.~T. Patera,
\newblock Calcolo {\bf 46}, 157 (2009).

\bibitem{Chen:2010:CRB:1958598.1958625}
Y.~Chen, J.~S. Hesthaven, Y.~Maday, and J.~Rodr\'{\i}guez,
\newblock SIAM J. Sci. Comput. {\bf 32}, 970 (2010).

\bibitem{Knezevic20111455}
D.~J. Knezevic and J.~W. Peterson,
\newblock Computer Methods in Applied Mechanics and Engineering {\bf 200}, 1455
   (2011).

\bibitem{Quarteroni}
A.~Quarteroni, G.~Rozza, and A.~Manzoni,
\newblock Journal of Mathematics in Industry {\bf 1}, 1 (2011).

\bibitem{Field:2011mf}
S.~E. Field {\em et~al.},
\newblock Phys. Rev.Lett. {\bf 106}, 221102 (2011), arXiv:1101.3765.

\bibitem{Caudill:2011kv}
S.~Caudill, S.~E. Field, C.~R. Galley, F.~Herrmann, and M.~Tiglio,
\newblock Class. Quant. Grav. {\bf 29}, 095016 (2012), arXiv:1109.5642.
%%CITATION = ARXIV:1109.5642;%%

\bibitem{PhysRevD.86.084046}
S.~E. Field, C.~R. Galley, and E.~Ochsner,
\newblock Phys. Rev. D {\bf 86}, 084046 (2012).

\bibitem{Cannon:2011xk}
K.~Cannon, C.~Hanna, and D.~Keppel,
\newblock Phys.Rev. {\bf D84}, 084003 (2011), arXiv:1101.4939.
%%CITATION = ARXIV:1101.4939;%%

\bibitem{Binev10convergencerates}
P.~Binev {\em et~al.},
\newblock SIAM J. Math. Analysis {\bf 43}, 1457 (2011).

\bibitem{DeVore2012}
R.~DeVore, G.~Petrova, and P.~Wojtaszczyk,
\newblock Arxiv preprint arXiv:1204.2290  (2012).

\bibitem{Cannon:poster}
D.~Keppel, K.~Cannon, M.~Frei, and C.~Hanna,
\newblock http://www.gravity.phys.uwm.edu/conferences/gwpaw/posters/keppel.pdf.

\bibitem{Cannon:2010qh}
K.~Cannon {\em et~al.},
\newblock Phys. Rev. {\bf D82}, 044025 (2010), arXiv:1005.0012.

\bibitem{Press92}
W.~H. Press, B.~P. Flannery, S.~A. Teukolsky, and W.~T. Vetterling,
\newblock {\em Numerical Recipes}, 2nd ed. (Cambridge University Press, New
  York, 1992).

\bibitem{Quarteroni2010}
A.~Quarteroni, R.~Sacco, and F.~Saleri,
\newblock {\em Numerical Mathematics} (Springer, Berlin, 2010).

\bibitem{chaturantabut:2737}
S.~Chaturantabut and D.~C. Sorensen,
\newblock SIAM Journal on Scientific Computing {\bf 32}, 2737 (2010).

\bibitem{Chaturantabut5400045}
S.~Chaturantabut and D.~Sorensen,
\newblock Discrete empirical interpolation for nonlinear model reduction,
\newblock in {\em Decision and Control, 2009 held jointly with the 2009 28th
  Chinese Control Conference. CDC/CCC 2009. Proceedings of the 48th IEEE
  Conference on}, pp. 4316 --4321, 2009.

\bibitem{Eftang:2011}
J.~L. Eftang and B.~Stamm,
\newblock International Journal for Numerical Methods in Engineering {\bf 90},
  412 (2012).

\bibitem{Aanonsen2009}
T.~O. Aanonsen,
\newblock {\em Empirical Interpolation with Application to Reduced Basis
  Approximations},
\newblock PhD thesis, Norwegian University of Science and Technology, 2009.

\bibitem{daCostaRibeiro:2011:CSS:2186225.2186298}
S.~da~Costa~Ribeiro, M.~Kleinsteuber, A.~M\"{o}ller, and M.~Kranz,
\newblock A compressive sensing scheme of frequency sparse signals for mobile
  and wearable platforms,
\newblock in {\em Proceedings of the 13th international conference on Computer
  Aided Systems Theory - Volume Part II}, EUROCAST'11, pp. 510--518, Berlin,
  Heidelberg, 2012, Springer-Verlag.

\bibitem{annealing2}
N.~Metropolis, A.~W. Rosenbluth, M.~N. Rosenbluth, A.~H. Teller, and E.~Teller,
\newblock The Journal of Chemical Physics {\bf 21}, 1087 (1953).

\bibitem{Galley:2010rc}
C.~R. Galley, F.~Herrmann, J.~Silberholz, M.~Tiglio, and G.~Guerberoff,
\newblock Class. Quantum Grav. {\bf 27}, 245007 (2010), arXiv:1005.5560.

\end{thebibliography}

\end{document}